\documentclass[preprint,12pt]{elsarticle}

\makeatletter
\def\ps@pprintTitle{%
  \def\@oddhead{}%
  \def\@evenhead{}%
  \def\@oddfoot{}%
  \def\@evenfoot{}%
}
\makeatother

\usepackage{amssymb}
\usepackage{amsmath}
\usepackage{xcolor}%
\usepackage{comment}
\usepackage{hyperref}
\usepackage[notcite,notref]{}

\newtheorem{example}{Example}
\newtheorem{definition}{Definition}
\newtheorem{axiom}{Axiom}
\newtheorem{lemma}{Lemma}

\newtheorem{proposition}{Proposition}

\newtheorem{theorem}{Theorem}
\newtheorem{proofinner}{Proof of Lemma}
\newenvironment{proof of lemma}[1][]
 {\if\relax\detokenize{#1}\relax
    \renewcommand\theproofinner{\thetheorem}%
  \else
    \renewcommand{\theproofinner}{#1}%
  \fi
  \proofinner}
 {\endproofinner}

\newtheorem{proofinnerP}{Proof of Proposition}
\newenvironment{proof of proposition}[1][]
 {\if\relax\detokenize{#1}\relax
    \renewcommand\theproofinnerP{\thetheorem}%
  \else
    \renewcommand{\theproofinnerP}{#1}%
  \fi
  \proofinnerP}
 {\endproofinnerP}

 \newtheorem{proofinnerT}{Proof of Theorem}
\newenvironment{proof of theorem}[1][]
 {\if\relax\detokenize{#1}\relax
    \renewcommand\theproofinnerT{\thetheorem}%
  \else
    \renewcommand{\theproofinnerT}{#1}%
  \fi
  \proofinnerT}
 {\endproofinnerT}
\newcommand{\cc}{c}
\newcommand{\sm}[1]{\textcolor{purple}{#1}}

\definecolor{pisa}{RGB}{12,102,124}
\newcommand\mic[1]{\textcolor{pisa}{#1}}

\begin{document}

\begin{frontmatter}



\title{Ranking alternatives from opinions on criteria}

 \author[label1]{Takahiro Suzuki\textsuperscript{[0000-0002-3436-6831]}}
   \author[label2]{Stefano Moretti\textsuperscript{[0000-0003-3627-3257]}}
   \author[label3]{Michele Aleandri\textsuperscript{[0000-0002-5177-8176]}}
 \affiliation[label1]{organization={Department of Civil Engineering, 
 Graduate School of Engineering, The University of Tokyo},
             addressline={Hongo Campus, 7-3-1 Hongo, Bunkyo-ku}, 
            city={Tokyo},
            postcode={113-8656}, 
            country={Japan}}

\affiliation[label2]
{organization={LAMSADE, CNRS, Universit{\'e} Paris-Dauphine, Universit{\'e} PSL},
             city={Paris},
             postcode={75016},
             country={France} \\
stefano.moretti@lamsade.dauphine.fr}

\affiliation[label3]{organization={LUISS University},
             addressline={Viale Romania, 32},
             city={Rome},
             postcode={00197},
             country={Italy} \\
             maleandri@luiss.it}

\begin{abstract}
A primary challenge in collective decision-making is that achieving unanimous agreement is difficult, even at the level of criteria. The history of social choice theory illustrates this: numerous normative criteria on voting rules have been proposed; however, disagreements persist regarding which criteria should take precedence. This study addresses the problem of ranking alternatives based on the aggregation of opinions over criteria that the alternatives might fulfill. Using the opinion aggregation model, we propose a new rule, termed the Intersection Initial Segment (IIS) rule, and characterize it using five axioms:  neutrality, independence of the worst set, independence of the best set, weak intersection very important player, and independence of non-unanimous improvement. We illustrate our approach on a running example where the objective is to rank voting rules, showing that our opinion aggregation model is particularly well-suited to this context, and that the IIS rule is a counterpart to the method discussed in Nurmi's paper (2015).
\end{abstract}



\end{frontmatter}

\section{Introduction}
\label{sec1}

\subsection{Overview of this research}
\label{sec1.1}
A key challenge in collective decision-making is the lack of unanimous agreement, even regarding normative axioms/criteria, as demonstrated by the history of social choice theory. Ever since the pioneering work by \cite{Arrow1951}  (for a comprehensive survey, see \cite{Brams2002,Brandt2016}), a number of normative axioms have been studied in social choice theory. Nevertheless, selecting an appropriate voting rule for a given
context remains a challenge: even among voting specialists,
 no consensus has been reached on an optimal voting procedure \cite{Laslier2011}. To this purpose, Nurmi \cite{Nurmi2015} presents an interesting problem:  how can an appropriate voting rule be selected based on the aggregation of voters' preferences over the properties they satisfy? This issue differs from the ordinary voting problem because voters do not express preferences over selected alternatives (i.e. voting rules). Instead, they are expected to have preferences over the criteria, each of which a voting rule either satisfies or does not. 
Nurmi informally proposes two possible methods for this problem.

This study aims to establish a general framework to select appropriate elements or alternatives (e.g., voting rules) based on people's opinions (e.g., preferences) on criteria (e.g., properties of voting rules). Our main result is the axiomatic characterization of a new opinion aggregator, termed \textit{intersection initial segment} (IIS) rule. It is an opinion aggregator that evaluates the elements by the intersection  of subsets of elements with the highest support. We prove that IIS is the unique opinion aggregator satisfying the following five fundamental axioms:
\begin{itemize}
\item The first axiom, named {\it NeuTrality} (NT), asserts that the  labels of elements do not matter in the evaluation. 
\item The second axiom, named  {\it Independence of the Worst Set} (IWS), states that if an opinion aggregator ranks an alternative strictly higher than another based on a support relation over subsets of elements, this ranking should be maintained in any situation where the support relation  differs from the original one only with respect to the structure of the worst equivalence class. 
\item The third axiom, named {\it Independence of the Best Set} (IBS), is the counterpart to IWS. It  focuses on the independence of strict rankings based on the partitions of the best equivalence class rather than the worst class. While IWS awards alternatives in numerous subsets with the highest support, the IBS penalizes the alternatives present in many subsets with the lowest support. 
\item The fourth axiom, named {\it Weak Intersection Very Important Player} (WIVIP), applies to situations where the quotient order of a support relation has only two classes. It requires that an alternative included in all subsets of the best equivalence class must be ranked strictly higher than any other alternative that does not exhibit the same behavior. 
\item the fifth and last axiom, named {\it Independence of Non-Unanimous Improvement} (INUI), asserts that the final ranking between two alternatives should remain unchanged when the support for subsets in a collection within the same equivalence class (where neither alternative is universally present) is slightly improved.
\end{itemize}

The scope of opinion aggregation is wide; it was proposed as an extended framework of social choice theory in which no specific reference to agents holding the opinions is needed \cite{Barbera2023a}, and it is also closely related to social ranking problem \cite{Suzuki}. In the present paper, the choice of voting rules is argued as a running example. Two opinion aggregators, IIS rule and \textit{support rule} \cite{Suzuki}, are shown to induce Nurmi's methods (Theorem~\ref{theo2}). Thus, our results suceed in giving an axiomatic grounds for the choice of voting rules (Nurmi's methods). 

\subsection{Related literature}
Methodologically, our analysis is based on the opinion aggregation model
\cite{Barbera2023a,Suzuki,Suzuki2024}. For a finite set $X$ of alternatives, the model assumes that there are opinions of the form "$S$ is at least as good as $T$", where $S,T\subseteq X$ are subsets of alternatives (\cite{Barbera2023a} originally considers opinions only on single alternatives; later, \cite{Suzuki,Suzuki2024a} generalized it so that opinions on sets of alternatives are considered), and an opinion aggregator determines a ranking of alternatives based on a list of opinions. In addition to the majority rule and Borda count considered in \cite{Barbera2023a},  some opinion aggregators exist in the literature: variants of majority rules \cite{Suzuki2024a}, and \textit{lexicographic support rule} (in \cite{Suzuki2024a}, it is shown to be a counterpart of the \textit{lexicographic excellence solution} (\textit{lex-cel}) of social ranking problem \cite{Bernardi2019}). Our IIS is a novel opinion aggregator (despite the apparent resemblance, it is different from lex-cel as shown in the discussion following Theorem \ref{theo1}). 

Regarding the choice of voting rules, while the axiomatic analysis of voting rules presents a promising approach for identifying a good voting rule, another stream of research considers the perspective that a good rule is one that aligns with voters preferences (referred to as procedural
autonomy \cite{Dietrich2005}). Within this framework, a typical assumption is that
voters are \textit{consequential}, meaning that they evaluate voting rules based
on the expected outcomes \cite{Barbera2004,Koray2006,Kultti2009,Rae1969,Suzuki2023}. Recent studies have examined
a procedural choice model that allows voters to have preferences, independent of these outcomes \cite{Houy2004,Nurmi2015,Suzuki2020a}.
Nurmi’s  proposal \cite{Nurmi2015} assumes that voters have preferences over the criteria (properties) that align well with the accumulation of axiomatic analyses in social choice theory, allowing  voting rules to be examined from the perspective of the  satisfied criteria.\\

\noindent
The remainder of the paper is organized as follows. Section \ref{sec2} presents preliminary notions and notations, and a description of the two Nurmi's methods studied in this paper. Section \ref{sec2.3} introduces an opinion aggregation model and its connections to a multi-criteria setting. Section \ref{sec:axioms} discusses five axioms of opinion aggregators.  Section \ref{sec3.2} showcases the most significant result of this study, that is, the axiomatic characterization of the IIS rule, using the axioms presented in the preceding section, and Section \ref{sec3.3} shows their logical independence. The connection between Nurmi's method and opinion aggregators is discussed in Section \ref{sec3}. Finally, Section \ref{sec4} concludes the paper with some remarks and new research directions.

\section{Model}
\label{sec2}


A binary relation $R$ on a set $A$ is called a \textit{weak order} if
it is reflexive, complete, and transitive. A weak order that is also
asymmetric is referred to as \textit{linear order}. The set of all weak orders
(resp.\ linear orders) on $A$ is denoted as $\mathcal{R}(A)$
(resp.\ $\mathcal{L}(A)$). For $R\in \mathcal{R}(A)$, we represent by 
$R$: $\Sigma_{1} \succ \cdots \succ \Sigma_{l}$, the fact that (i) the
$k^{\mathrm{th}}$ best equivalence class is $\Sigma_{k}$ (i.e., 
$\Sigma_{k} = \{ a \in A  \setminus 
( \Sigma_{1} \cup \cdots \cup \Sigma_{k - 1} )$: 
$(a,b)\in R,\ \forall b \in A \setminus 
(\Sigma_{1} \cup \cdots \cup \Sigma_{k - 1} ) \}$),
and (ii) if exactly $l$ equivalence classes exist, $R:\Sigma_{1} \succ \cdots \succ \Sigma_{l}$ is called
the \textit{quotient order} (q.o.) of $R$. For
$R\in \mathcal{R}(A)$, we denote the asymmetric and symmetric
parts as $P(R)$ and $I(R)$, respectively, where 
$P(R) := \{ (a,b)\colon (a,b) \in R$  and $(b,a) \notin R \}$
and $I(R) := \{ (a,b)\colon (a,b) \in R$  and $(b,a) \in R \}$. 
 For any set $A$, let $1_{A}$ be
the identity function on $A$, i.e., $1_{A}$: $A \rightarrow A$ such
that for all $a \in A$, $1_{A}(a) = a$. For a nonnegative integer,
$a$, let $\lbrack a\rbrack$ denote the set of all positive integers,
which is equal to or less than $a$; i.e. ,
$\lbrack a\rbrack := \{ n\mathbb{\in N : }1 \leq n \leq a \}$.

Let $X$ be a finite set of alternatives (voting rules in the context of procedural choices). We assume
that $3 \leq |X|$. Let
$\mathfrak{X } := 2^{X} \setminus \{ \emptyset \}$ denote the
set of all nonempty subsets of $X$. For $x \in X$ and
$\Delta\subseteq \mathfrak{X}$, we denote the set of all elements in
$\Delta$ that contain $x$ by 
$\Delta\lbrack x\rbrack := \{ S \in \Delta\colon x \in S \}$. We
denote the intersection of all the elements in $\Delta$ as
$\bigcap \Delta := \{ x \in X\colon x \in S,\ \forall S \in \Delta \}$.

Let $C = \{ 1,2,\cdots,m \}$ be a finite set of criteria.
Each alternative either satisfies or fails to satisfy each criterion.
We identify each criterion $c \in C$ with a 
corresponding criterion map $\cc:X \rightarrow$ $\{$True, False$\}$ such that for each alternative $x \in X$, $\cc(x)=\text{True}$ means that $x$ satisfies $c$, and $\cc(x)=\text{False}$  means that $x$ does not satisfy $c$.
For every $c \in C$, we
denote the set of all alternatives satisfying criterion $c$ as
$\mathrm{Tr}(c) := \{ x \in X$: $c(x) = \mbox{True}\}$. Additionally, let
$\mathrm{Tr}(C) := \{ \mathrm{Tr}(c)$: $c \in C \}$. We assume that no two
criteria in $C$ are logically equivalent with respect to $X$. That is,
for any $c,d \in C$,  $c \neq d \Rightarrow \mathrm{Tr}(c) 
\neq \mathrm{Tr}(d)$. We
assume that for each $c \in C$, there exists at least one
$x \in X$ such that $c(x)$ = True (no self-contradicting
criterion exists). Therefore, $\mathrm{Tr}$ is an injection from $C$ into 
$\mathfrak{X}$. For $S \in \mathfrak{X}$, let $\mathrm{Tr}^{*}(S)$
represent criterion $c$ such that $\mathrm{Tr}(c) = S$ (if $c$
exists), and $\mathrm{Tr}^{*}(S) = \emptyset$ if there is no such $c$ (if there exists $c$ such that $Tr(c)=S$, then such $c$ should be unique, because 
$\mathrm{Tr}$: $C\rightarrow \mathfrak{X}$ is an injection). Thus,
$\mathrm{Tr}^{*}:\mathfrak{X \rightarrow}C \cup \{ \emptyset \}$ are
well-defined. 
Furthermore, for all
$S \in  \mathrm{Tr}(C),$ it holds $ 
 \mathrm{Tr} \circ \mathrm{Tr}^*(S)=S$.

\begin{example}\label{ex:1}
To illustrate the proposed framework, we consider a shorter version of the example provided in \cite{Nurmi2015}.
A set of alternatives is formed using seven voting rules 
($X=\{Copeland,$  
$Dodgson, Maximin, Kemeny, Plurality,$ $ Borda,$ $ Approval\}$), and a set of six criteria $C=\{a,b,c,d,e,f\}$ 
is formed by the following six properties of voting rules:
$a$) the Condorcet winner criterion;
$b$) the Condorcet loser criterion;
$c$) the strong Condorcet criterion;
$d$) monotonicity;
$e$) Pareto;
$f$) consistency.

In Table \ref{tab:comparison}, which corresponds to a submatrix of Table 5 in \cite{Nurmi2015}, indicates whether voting rule $x \in X$, represented by a row, satisfies ($c(x)=$ True, denoted by T) or does not satisfy ($c(x)=$False, denoted by F) the criterion corresponding to a column.
\begin{table}[h!]
\centering
\begin{tabular}{|l|c|c|c|c|c|c|}
\hline
voting rule\ {\it$/$criteria} & \textit{a} & \textit{b} & \textit{c} & \textit{d} & \textit{e} & \textit{f} \\ \hline 
Copeland           & T          & T          & T          & T          & T          & F      \\ \hline 
Dodgson            & T          & F          & T          & F          & T          & F      \\ \hline 
Maximin            & T          & F          & T          & T          & T          & F     \\ \hline 
Kemeny             & T          & T          & T          & T          & T          & F    \\ \hline 
Plurality          & F          & F          & T          & T          & T          & T    \\ \hline 
Borda              & F          & T          & F          & T          & T          & T     \\ \hline 
Approval           & F          & F          & F          & T          & F          & T       \\ \hline 
\end{tabular}
\caption{The function $\cc:X \rightarrow $\{True, False\}, indicates which voting rules (alternatives) $x \in X$ (on the rows) satisfy (value T stands for True) or do not satisfy (value F stands for False) the properties (criteria) $c \in C$ (on the columns).  For more details on voting rules and the properties they satisfy, see \cite{Brandt2016} .}
\label{tab:comparison}
\end{table}
Using the previously introduced notation, we have:
\[
\begin{array}{l}
   \mathrm{Tr}(a) = \{Copeland,  Dodgson, Maximin, Kemeny \},\\
   \mathrm{Tr}(b) = \{Copeland,  Kemeny, Borda \}, \\
\mathrm{Tr}(c) = \{Copeland,  Dodgson, Maximin, Kemeny, Plurality \},\\
\mathrm{Tr}(d) = \{Copeland, Maximin, Kemeny, Plurality, Borda, Approval \},\\
\mathrm{Tr}(e) = \{Copeland, Dodgson, Maximin, Kemeny, Plurality, Borda \},\\
\mathrm{Tr}(f) = \{ Plurality, Borda, Approval \}.
\end{array}
\]
Moreover, $\mathrm{Tr}^{*}(S)=\emptyset$ for all $S \in \mathfrak{X} \setminus \mathrm{Tr}(C)$ where $\mathrm{Tr}(C)=\{\mathrm{Tr}(a), \ldots, \mathrm{Tr}(f)\}$.

\end{example}

Let $N = \{ 1,2,\cdots,n \}$ be the set of voters. 
An $n$-tuple $\succsim$ $=$ $(\succsim_{1},\cdots, \succsim_{n})$ 
$\in$ $\mathcal{L}(C)^{n}$ of linear orders on $C$ 
is called a \textit{preference profile}. 
For each $i\in N$, the linear order $\succsim_{i}$ is interpreted as voter
$i$'s preference over $C$.

\begin{definition}
\label{def1}\normalfont%

A {\it Criteria-based Social Choice Correspondence} (\textit{CSCC}) is a function
$F\colon\mathcal{L}(C)^{n} \rightarrow \mathfrak{X}$.

\end{definition}


We now introduce two concepts of CSCC from \cite{Nurmi2015}.

\subsubsection*{Nurmi's first method $F^{N1}$.}

The first method consists of two steps. First, Borda
count\footnote{Specifically, Nurmi (2015) \cite{Nurmi2015} discusses the application of  
Kemeny's rule and the Borda count in this step, arguing that ``both 
rules resort to the same metric but with different end states. In this context, Kemeny's rule appears to be more appropriate 
because the choice procedure uses the following performance: 
table (p.247).'' However, in this study, we assume 
the Borda count is used in this step and its 
justification is provided later.} is applied to yield a weak order 
$B_{\succsim}$ on the criteria. Subsequently, the choice set
$F^{N1}( \succsim ) \in \mathfrak{X}$ is determined by analyzing the criteria
lexicographically. For instance, assume that
$B_{\succsim}\colon 1 \succ 2 \succ \cdots \succ m$. 
(i.e., criterion $k$ is the $k^{\mathrm{th}}$ best). 
First, the method focuses on $T_{1} := \mathrm{Tr}(1)$. 
If $T_{1}$ is a singleton, $T_{1}$ is selected. If $T_{1}$
contains multiple alternatives, the method focuses on
$T_{2} := \mathrm{Tr}(1) \cap \mathrm{Tr}(2)$. 
If $T_{2}$ is empty, then
$T_{1}$ is selected. Otherwise, as in the previous step, if $T_{2}$ is 
 singleton, $T_{2}$ is
selected. If $T_{2}$ contains multiple alternatives, then the method
focuses on $T_{3} := \mathrm{Tr}(1) \cap \mathrm{Tr}(2) 
\cap \mathrm{Tr}(3)$. 
In this manner, if (i) at least one alternative satisfies all 
criteria $1,2,\cdots,(k - 1)$ but (ii) no alternatives satisfy all the criteria
$1,2,\cdots,k$ then $F^{N1}( \succsim )$ is determined by
$\mathrm{Tr}(1) \cap \cdots \cap \mathrm{Tr}(k - 1)$. 
This method denoted as $F^{N1}$ is referred to as \textit{Nurmi's first method}. To provide a formal definitions, we first need to introduce some further notation.

For any $ \succsim \in \mathcal{ L}(C)^{n}$, $c \in C$ and $x \in X$,
we define $s_{\succsim}^{B}(c)$ (the \textit{Borda score of the criterion}
$c$ at $\succsim$) as
\begin{align*}
 s_{\succsim}^{B}(c)
  := \sum_{i \in N} | \{ d \in C\colon c \succsim_{i}d \} |.
\end{align*}

The associate weak order 
$B_{\succsim}\in \mathcal{R}(C)$ for all $c,c' \in C$ is such that $$( c,c' ) \in B_{\succsim} \Leftrightarrow s_{\succsim}^{B}(c), 
\geq s_{\succsim}^{B}( c' ).$$

\begin{definition}[Nurmi's first method $\boldsymbol{{F}^{N1}}$]
\label{def2}\normalfont%
For $\succsim = ( \succsim_{1},\cdots, \succsim_{n} ) \in$\linebreak 
$\mathcal{ L}(C)^{n}$,
let $B_{\succsim}$ be $\Sigma_{1} \succ \cdots \succ \Sigma_{l}$. For each
$k = 1,\cdots,l$, let $T_{k} := \{ x \in X$: $\forall k' \leq k, 
\forall c \in \Sigma_{k'},x \in \mathrm{Tr}(c) \}$.
We define $F^{N1}( \succsim )$ as follows%
\footnote{To the best of our knowledge, Nurmi (2015) 
does not specify $F( \succsim )$ when no alternative satisfies all 
top-ranked criteria according to collective ranking $B_{\succsim}$. 
In this case, we assume $F^{N1}( \succsim ) = X$.}.
\begin{align*}
F^{N1}( \succsim ) 
= \left\{ 
 \begin{array}{@{}ll@{}}
  X  & \mathrm{if}\   T_{1} = \emptyset, \\
  T_{k - 1}  & \mathrm{if}\  T_{k - 1} \neq \emptyset = T_{k}\ (2 \leq k \leq l), \\
  T_{l} & \mathrm{if}\   T_{l} \neq \emptyset.
\end{array} \right.
\end{align*}
\end{definition}

\subsubsection*{Nurmi's second method $F^{N2}$.}

The second method estimates the score for each voting
rule by summing the Borda scores for the criteria that satisfies the rule; see page 248 in \cite{Nurmi2015}. For $\succsim \in \mathcal{ L}(C)^{n}$ and
$x \in X$, we define $s_{\succsim}^{B}(x)$ (the \textit{Borda score of
alternative} $x$ at $\succsim$) as follows:
\begin{align*}
\hat{s}_{\succsim}^{B}(x) 
 := \sum_{c \in C:\, \mathrm{Tr}(c)\ni x}
 s_{\succsim}^{B}(c).
\end{align*}

\begin{definition}[Nurmi's second method $\boldsymbol{F^{N2}}$]
\label{def3}\normalfont
For all $\succsim \in \mathcal{ L}(C)^{n}$ in \textit{Nurmi's second method}, 
$F^{N2}:\mathcal{L}(C)^n\to\mathfrak{X}$ is defined as follows:
\begin{equation*}
    F^{N2}( \succsim ) := \underset{x \in X} {\arg\max}\, \hat{s}_{\succsim}^{B}(x).
\end{equation*}
\end{definition}

\begin{example}\label{ex:2}
Consider a set of alternatives $X,$ along with a set of criteria $C$ and the criterion maps $\cc$ for each $c \in C$, as described in Example \ref{ex:1}.
Additionally, consider a set of three voters $N=\{1,2,3\}$ whose preference profiles over the criteria in $C$ are such that:
\begin{eqnarray*}
& &\succsim_1: a \succ_1 b \succ_1 c \succ_1 d \succ_1 e   \succ_1 f; \\
& &\succsim_2: d \succ_2 c \succ_2 b \succ_2  a \succ_2 f   \succ_2 e; \\
& &\succsim_3: f \succ_3 e \succ_3 d \succ_3  c \succ_3 b   \succ_3 a. \\
\end{eqnarray*}
We obtain the following Borda scores for the criteria:
\[
\begin{array}{l}
s_{\succsim}^{B}(a)=6+3+1=10,\\
s_{\succsim}^{B}(b)=5+4+2=11,\\
s_{\succsim}^{B}(c)=4+5+3=12,\\
s_{\succsim}^{B}(d)=3+6+4=13,\\
s_{\succsim}^{B}(e)=2+1+5=8,\\
s_{\succsim}^{B}(f)=1+2+6=9.
\end{array}
\]
Thus, the Borda ranking of criterion $B_\succsim$ is such that:
\[
B_\succsim: d \succ c \succ b \succ a \succ f \succ e.
\]
According to Definition \ref{def2}, we obtain
\[
\begin{array}{l}
T_1=\mathrm{Tr}(d) = \{Copeland, Maximin, Kemeny, Plurality, Borda, Approval \},\\
T_2 = \{Copeland,   Maximin, Kemeny, Plurality \},\\
T_3 = \{Copeland,  Kemeny \}, \\
T_4 = \{Copeland,  Kemeny \},\\
T_5 = \emptyset.
\end{array}
\]
Therefore, Nurmi's first method $\boldsymbol{{F}^{N1}}$ yields set $T_4 = \{Copeland,  Kemeny \}$.

We now compute the following Borda scores for the alternatives:
\[
\begin{array}{l}
\hat{s}_{\succsim}^{B}(Copeland)=10+11+12+13+8=54,\\
\hat{s}_{\succsim}^{B}(Dodgson)=10+12+8=30,\\
\hat{s}_{\succsim}^{B}(Maximin)=10+12+13+8=43,\\
\hat{s}_{\succsim}^{B}(Kemeny)=10+11+12+13+8=54,\\
\hat{s}_{\succsim}^{B}(Plurality)=12+13+8+9=42,\\
\hat{s}_{\succsim}^{B}(Borda)=11+13+8+9=41,\\
\hat{s}_{\succsim}^{B}(Approval)=13+9=22.
\end{array}
\]
In this example, Nurmi's second method $\boldsymbol{{F}^{N2}}$ also yields set $\{Copeland,$ $  Kemeny \}$.
\end{example}

\section{Opinion aggregation}
\label{sec2.3}

An opinion aggregation model has been recently proposed by \cite{Barbera2023a,Barbera2023b} to address
preferences over opinions that are not necessarily transitive (or even
acyclic). The model has been extended to opinions on sets of
alternatives in \cite{Suzuki,Suzuki2024} and  our basic notation follows that of \cite{Suzuki}.

A function $o$: $\mathfrak{X} \times \mathfrak{X} \rightarrow 
\mathbb{Z}_{\geq 0}$ that associates to any ordered pair of subsets of alternatives in $X$ a nonnegative integer number, is called a \textit{state of opinion}. 
For every $S,T \in \mathfrak{X}$, $o(S,T)$ represents the number of opinions
according to which $S$ is at least as good as $T$. 
We denote 
$\mathcal{O}( \mathfrak{X} )$ as the set of all states of
opinion of $\mathfrak{X}$. An \textit{opinion aggregator} is a function
$f:\mathcal{O}( \mathfrak{X}) \rightarrow \mathcal{R}(X)$, associating to any state of opinion in $\mathcal{O}( \mathfrak{X} )$ a weak order over the alternatives in $X$. 
In most previous studies of opinion aggregation \cite{Barbera2023a,Suzuki,Suzuki2024}, 
the codomain of the opinion aggregator is assumed to be
 a set of all (complete and reflexive) binary relationships. In this study, we assume that the codomain of the opinion aggregator is the
set of all weak orders in $X$. 

For any $o \in \mathcal{O}( \mathfrak{X} )$ and every $S \in \mathfrak{X}$, we define the \textit{support} of $S$ as 
\[
m_{o}(S) := \sum_{T \in \mathfrak{X}} o(S,T).
\]
We denote by $M_{o}\subseteq \mathfrak{X} \times \mathfrak{X}$ the \emph{support relation} on $\mathfrak{X}$, such that $$(S,T) \in M_{o} 
\Leftrightarrow m_{o}(S) \geq m_{o}(T),$$ and we denote its quotient order as
$M_{o}:\Sigma_{1} \succ \cdots \succ \Sigma_{l}$, where each equivalence class $\Sigma_{k}$ contains subsets of $X$ that share the same support and have greater support than those in $\Sigma_{k+1}$ for all $k \in \{1, \ldots, l-1\}$. We emphasize that any vector of nonnegative integers $v \in \mathbb{Z}_{\geq 0}^\mathfrak{X}$ can be obtained as a support of some state of opinion $o \in \mathcal{O}( \mathfrak{X} )$ (for instance, consider $o \in \mathcal{O}( \mathfrak{X} )$ such that for each $S \in \mathfrak{X}$ there exists $T_S \in \mathfrak{X}$ with $o(S,T_S)=v_S$ and $o(S,T)=0$ for every $T \in \mathfrak{X}$ with $T \neq T_S).$

To introduce two opinion aggregators, the {\it IIS rule} and {\it support rule}, we first require further preliminary notation. For any $o \in \mathcal{O}( \mathfrak{X} )$ and the corresponding support relation $M_o:\Sigma_{1} \succ \cdots \succ \Sigma_{l}$, let
$$T_{k}' := \bigcap  ( \bigcup_{k' \leq k} \Sigma_{k'} ) = \{ x \in X: \forall k' \leq k,\forall S \in \Sigma_{k'},x \in S \},$$ and, for every $x \in X$,
let $e_{o}(x)$ be an integer such that 
$$ e_{o}(x)= \max\{ k \in \lbrack l\rbrack\colon x \in T_{k}' \},$$ if $k$ exists, and  $e_{o}(x)=0$ otherwise.

\begin{definition}[opinion aggregators]
\label{def4}\normalfont%
For any $x,y \in X$ 
and $o \in \mathcal{O}( \mathfrak{X} )$, \textit{IIS rule} $f^{IIS}$, and 
\textit{support rule} $f^{S}$ are defined as follows:
\begin{itemize}
\item[-]
  $(x,y) \in f^{IIS}(o) \Leftrightarrow e_{o}(x) \geq e_{o}(y)$.
\item[-]
  $(x,y) \in f^{S}(o) \Leftrightarrow \sum_{T \in \mathfrak{X}\lbrack x\rbrack}
  m_{o}(T) \geq \sum_{T \in \mathfrak{X}\lbrack y\rbrack} m_{o}(T)$.
\end{itemize}
\end{definition}

A state of opinion can be generated in various ways, such as by deriving it from individual preferences over criteria. Thus, a criterion
$c$ is identified by $\mathrm{Tr}(c)$, and the preference profile is
interpreted as a collection of opinions on $\mathfrak{X}$. We  introduce the following definitions.
\begin{definition}[induced state of opinion]\footnote{This translation of an $n$-tuple of binary relations into a state 
of opinion extends the argument presented in \cite{Suzuki} for $n = 1$.}
\label{def6}\normalfont%
Let
$\succsim = \left( \succsim_{1},\cdots, \succsim_{n} \right) \in$
\linebreak 
$\mathcal{ L}(C)^{n}$. 
The \textit{induced state of opinion} (from $\succsim$), 
denoted by $o_{\succsim}$, is defined, for any $S,T \in \mathfrak{X}$, as follows:
\[
o_{\succsim}(S,T) 
 := \left| \left\{ i \in N\colon  \mathrm{Tr}^{*}(S) \succsim_{i} 
 \mathrm{Tr}^{*}(T) \right\} \right|.
\]
\end{definition}

Note that $m_{o_\succsim}(S)=0$ for each $S \in \mathfrak{X} \setminus \mathrm{Tr}(C)$ because $o_\succsim(S,T)=|\emptyset|=0$, for any $T \in  \mathfrak{X}$.

The following proposition highlights some connections between the Borda scores considered in Section \ref{sec2} and the support for an induced state of opinion.

\begin{proposition}\label{prop:bordamo}
Let
$\succsim = \left( \succsim_{1},\cdots, \succsim_{n} \right) \in\mathcal{ L}(C)^{n}$ and let $o_{\succsim}$ be the
 induced state of opinion from $\succsim$. We obtain
 \begin{equation} \label{eq:eqm0}
      s_{\succsim}^{B}(c)=m_{o_{\succsim}}(\mathrm{Tr}(c)),
 \end{equation}
for every $c \in C$, and 
\[
\hat{s}_{\succsim}^{B}(x)=\sum_{S \in \mathfrak{X}\lbrack x\rbrack}
  m_{o_{\succsim}}(S),
\]
for every $x \in X$.
\end{proposition}
\begin{proof of proposition}[\ref{prop:bordamo}]
For every $c \in C$ we obtain 
\begin{eqnarray*}\label{eq:eqm0}
s_{\succsim}^{B}(c)&=&\sum_{i \in N}|\{d \in C: c \succsim_i d\}|\\ \nonumber
&=&\sum_{d \in C}|\{i \in N: c \succsim_i d\}|\\ \nonumber
&=&\sum_{d \in C}|\{i \in N: \mathrm{Tr}^{*}(\mathrm{Tr}(c)) \succsim_i  \mathrm{Tr}^{*}(\mathrm{Tr}(d))\}|\\ \nonumber
&=&\sum_{T \in \mathfrak{X}}|\{i \in N: \mathrm{Tr}^{*}(\mathrm{Tr}(c)) \succsim_i  \mathrm{Tr}^{*}(T)\}|\\ \nonumber
&=&\sum_{T \in \mathfrak{X}}o_{\succsim}(\mathrm{Tr}^{*}(\mathrm{Tr}(c)),  \mathrm{Tr}^{*}(T))\\ \nonumber
&=&m_{o_{\succsim}}(\mathrm{Tr}(c)).\\
\nonumber
\end{eqnarray*}
Moreover, for every $x \in X$ we obtain 
\begin{eqnarray}  \nonumber
\hat{s}_{\succsim}^{B}(x)&=&\sum_{c \in C:\, \mathrm{Tr}(c)\ni x} s_{\succsim}^{B}(c)\\ \nonumber
&=&\sum_{S \in \mathfrak{X}\lbrack x\rbrack}
  m_{o_{\succsim}}(S),\\ \nonumber
\end{eqnarray}
where the last equality follows from the fact that, by relation (\ref{eq:eqm0}),  $s_{\succsim}^{B}(\mathrm{Tr}^{*}(S))=m_{o_{\succsim}}(S)$, for every $S \in \mathrm{Tr}(C)$, and $m_{o_{\succsim}}(S)=0$, for every $S \in \mathfrak{X} \setminus \mathrm{Tr}(C)$. 
\end{proof of proposition}

\begin{example}\label{ex:3}
Additionally, consider $X, A, C, N$ and $(\succsim_i)_{i \in N}$ in Example \ref{ex:2}. Values of the induced state of opinion $o_{\succsim}(S,T)$ with $S,T \in \{\mathrm{Tr}(a), \ldots, \mathrm{Tr}(f)\}$ 
are presented in Table \ref{tab:opaggr}.
\begin{table}[h!]
    \centering
\begin{tabular}{|c|c|c|c|c|c|c|}
\hline 
 & $\mathrm{Tr}(a)$ & $\mathrm{Tr}(b)$ & $\mathrm{Tr}(c)$ & $\mathrm{Tr}(d)$ & $\mathrm{Tr}(e)$ & $\mathrm{Tr}(f)$ \\
\hline 
$\mathrm{Tr}(a)$ & 3 & 1 & 1 & 1 & 2  & 2\\
\hline 
$\mathrm{Tr}(b)$ & 2 & 3 & 1 & 1 & 2 & 2 \\
\hline 
$\mathrm{Tr}(c)$ & 2 & 2 & 3 & 1 & 2 & 2 \\
\hline 
$\mathrm{Tr}(d)$ & 2 & 2 & 2 & 3 & 2 & 2 \\
\hline 
$\mathrm{Tr}(e)$ & 1 & 1 & 1 & 1 & 3 & 1 \\
\hline 
$\mathrm{Tr}(f)$ & 1 & 1 & 1 & 1 & 2 & 3 \\
\hline
\end{tabular} 
    \caption{Values of $o_{\succsim}(S,T)$ for all $S,T \in \{\mathrm{Tr}(a), \ldots, \mathrm{Tr}(f)\}$ in the situation considered in Example \ref{ex:2}.}
    \label{tab:opaggr}
\end{table}


Note that $o_{\succsim}(S,T)=0$ for all other $S,T \in \mathfrak{X}$ such that either $S \notin \{\mathrm{Tr}(a), \ldots, \mathrm{Tr}(f)\}$ or $T \notin \{\mathrm{Tr}(a), \ldots, \mathrm{Tr}(f)\}$.

We find that the support $m_{o_{\succsim}}$  for the subsets of voting rules is such that
\[
\begin{array}{l}
m_{o_{\succsim}}(\mathrm{Tr}(a))=10,\\
m_{o_{\succsim}}(\mathrm{Tr}(b))=11,\\
m_{o_{\succsim}}(\mathrm{Tr}(c))=12,\\
m_{o_{\succsim}}(\mathrm{Tr}(d))=13,\\
m_{o_{\succsim}}(\mathrm{Tr}(e))=8,\\
m_{o_{\succsim}}(\mathrm{Tr}(f))=9,
\end{array}
\]
and $m_{o_{\succsim}}(S)=0$ for each $S \in \mathfrak{X} \setminus \{\mathrm{Tr}(a), \ldots, \mathrm{Tr}(f)\}$. 
Thus, we obtain the following values for $e_{o_{\succsim}}$:
\[
\begin{array}{l}
     e_{o_{\succsim}}(Copeland)= 4, \\
     e_{o_{\succsim}}(Dodgson)= 0, \\
     e_{o_{\succsim}}(Maximin)= 2, \\
     e_{o_{\succsim}}(Kemeny)= 4, \\
     e_{o_{\succsim}}(Plurality)= 2, \\
     e_{o_{\succsim}}(Borda)= 1, \\
     e_{o_{\succsim}}(Approval)= 1.
\end{array}
\]
The IIS rule $f^{IIS}$ yields the ranking 
$$f^{IIS}:Copeland, Kemeny \succ Maximin, Plurality \succ  Approval,  Borda \succ  Dodgson.\footnote{Brackets are omitted; so, for instance, $Copeland, Kemeny$ denotes the equivalence class $\{Copeland, Kemeny\}$.}$$

Moreover, we obtain the values of
\[
\begin{array}{l}
     \sum_{S \in \mathfrak{X}\lbrack Copeland\rbrack}
  m_{o_{\succsim}}(S)=10+11+12+13+8=54,\\
     \sum_{S \in \mathfrak{X}\lbrack Dodgson\rbrack}
  m_{o_{\succsim}}(S)= 10+12+8=30,\\
     \sum_{S \in \mathfrak{X}\lbrack Maximin\rbrack}
  m_{o_{\succsim}}(S)=10+12+13+8=43,\\
     \sum_{S \in \mathfrak{X}\lbrack Kemeny\rbrack}
  m_{o_{\succsim}}(S)=10+11+12+13+8=54,\\
     \sum_{S \in \mathfrak{X}\lbrack Plurality\rbrack}
  m_{o_{\succsim}}(S)=12+13+8+9=42,\\
     \sum_{S \in \mathfrak{X}\lbrack Borda\rbrack}
  m_{o_{\succsim}}(S)=11+13+8+9=41,\\
     \sum_{S \in \mathfrak{X}\lbrack Aproval\rbrack}
  m_{o_{\succsim}}(S)=13+9=22.\\
\end{array}
\]
Thus, the support rule $f^{S}$ yields the ranking 
$$f^{S}:Copeland, Kemeny \succ Maximin \succ Plurality \succ Borda \succ Dodgson \succ Approval.$$

\end{example}

\section{Axioms for opinion aggregators}\label{sec:axioms}

Consider a permutation $\pi:X\to X$ and an opinion $o\in\mathcal{O}(\mathfrak{X})$, then define the state of opinion $o^\pi\in\mathcal{O}(\mathfrak{X})$ as
\begin{equation*}
    o^\pi(S,T)=o(\pi(S),\pi(T)),\quad \forall S,T\in\mathfrak{X},
\end{equation*}
where $\pi(S)=\{\pi(x):\, x\in S \}$. The support of $o^\pi$ and the support relation are given by
\begin{align*}
    m_{o^\pi}(S)&=m_{o}\big(\pi(S)\big),\quad \forall S\in\mathfrak{X};\\
    (S,T)\in M_{o^\pi} &\iff \big(\pi(S),\pi(T)\big)\in M_{o},\quad \forall S,T\in\mathfrak{X}. 
\end{align*}

\begin{axiom}[Neutrality, NT] \label{NT}
An opinion aggregator $f:\mathcal{O}(\mathfrak{X})\rightarrow  \mathcal{R}(X)$
satisfies the   \textit{NT} axiom if, for any permutation $\pi$ on $X$ and $x,y\in X$, it holds that 
\[
 \left\lbrack (x,y) \in R ( f(o)) 
  \Leftrightarrow \big(\pi(x),\pi(y)\big) \in R \left( f\left( o^\pi \right) \right) \right\rbrack.
\]
\end{axiom}

Neutrality axioms is a standard property asked in literature and it requires that an opinion aggregator should not depend on the labels of the elements of $X$ (see, for instance, \cite{Bernardi2019, Suzuki2024c,  Suzuki}; with a similar meaning, also consider the anonymity axiom for solutions of cooperative games \cite{Shapley1953}).

\begin{axiom}[Independence of the Worst Set, IWS] \label{IWS}
An opinion aggregator $f:\mathcal{O}(\mathfrak{X})\rightarrow  \mathcal{R}(X)$ satisfies the \textit{IWS} axiom if, for every $x,y \in X$ and  $o^1, o^2 \in \mathcal{O}( \mathfrak{X} )$ such that the associated quotient orders are $M_{o^1}:\Sigma_{1} \succ \cdots \succ \Sigma_{l}$ and  $M_{o^2}:\Sigma_{1} \succ \cdots \succ \Sigma_{l - 1} 
\succ \Gamma_{1} \succ \cdots  \succ \Gamma_{l'}$ with $\Sigma_l=\Gamma_{1} \cup \cdots  \cup \Gamma_{l'}$, it holds that

\[
 \left\lbrack (x,y) \in P ( f(o^1)) 
  \Rightarrow (x,y) \in P \left( f\left( o^2 \right) \right) \right\rbrack.
\]

\end{axiom}

The IWS axiom states that if an opinion aggregator ranks an alternative $x$ strictly higher than an alternative $y$ in an induced state of opinion $o^1$, it must continue to rank $x$ strictly higher than $y$ in an induced state of opinion $o^2$ whose associated quotient order $M_{o^2}:\Sigma_{1} \succ \cdots \succ \Sigma_{l - 1} 
\succ \Gamma_{1} \succ \cdots  \succ \Gamma_{l'}$  differs from  that of $o^1$ with respect to subsets in classes $\Gamma_{1}, \cdots, \Gamma_{l'}$. These classes form a partition of the worst equivalence class $\Sigma_l$ according to $m_{o^1}$. The IWS axiom is designed to prioritize opinions that show strong support across multiple subset, and is also a property studied in \cite{Bernardi2019} for evaluating the excellence of individuals in the setting of social ranking.  
Note that such a situation occurs, for instance, when a new criterion that is worse than any other according to $\succsim$ is introduced into the analysis, as illustrated in the following example. 

\begin{example}\label{ex:iws}
Let $C'=C \setminus \{e\}$.
Consider $X, A, C', N$, $(\succsim'_i)_{i \in N}$ and $o_{\succsim'}$ as in Examples \ref{ex:2} and \ref{ex:3} after the removal of criterion $e$. It is easy to verify that the support $m_{o_{\succsim'}}$ of the subsets of the voting rules is such that
\[
\begin{array}{l}
m_{o_{\succsim'}}(\mathrm{Tr}(a))=8,\\
m_{o_{\succsim'}}(\mathrm{Tr}(b))=9,\\
m_{o_{\succsim'}}(\mathrm{Tr}(c))=10,\\
m_{o_{\succsim'}}(\mathrm{Tr}(d))=11,\\
m_{o_{\succsim'}}(\mathrm{Tr}(f))=7,
\end{array}
\]
and $m_{o_{\succsim'}}(S)=0$ for each $S \in \mathfrak{X} \setminus \{\mathrm{Tr}(a), \ldots, \mathrm{Tr}(d), \mathrm{Tr}(f)\}$. Therefore, the worst equivalence class according to $m_{o_{\succsim'}}$ is 
\[
\mathfrak{X} \setminus \{\mathrm{Tr}(a), \ldots, \mathrm{Tr}(d), \mathrm{Tr}(f)\}.
\]
Now, consider $X, A, C, N$, $(\succsim_i)_{i \in N}$ and $o_{\succsim}$ as in Examples \ref{ex:2} and \ref{ex:3}. As illustrated in Example \ref{ex:3}, the two worst equivalence classes according to $m_{o_\succsim}$ are such that
\[
\{\mathrm{Tr}(e)\} \succ \mathfrak{X} \setminus \{\mathrm{Tr}(a), \ldots, \mathrm{Tr}(f)\},
\]
Notice that $\{\{\mathrm{Tr}(e)\}, \mathfrak{X} \setminus \{\mathrm{Tr}(a), \ldots, \mathrm{Tr}(f)\}
$ forms a partition of $\mathfrak{X} \setminus$ $\{\mathrm{Tr}(a),\ldots,$ $ \mathrm{Tr}(d),$ $\mathrm{Tr}(f)\}$ and the subsets within $ \{\mathrm{Tr}(a), \ldots, \mathrm{Tr}(d), \mathrm{Tr}(f)\}$ are ranked in the same manner according to both $m_{o_{\succsim'}}$ and $m_{o_\succsim}$. The IWS axiom requires that if, for instance, an opinion aggregator ranks Kemeny  strictly higher than Plurality in $o_{\succsim'}$ (which corresponds to $o^1$ according to the notations used in Axiom \ref{IWS}), then the aggregator must also rank Kemeny strictly higher than Plurality in $o_{\succsim}$ (which corresponds to $o^2$ according to the notations used in Axiom \ref{IWS}).
\end{example}

\begin{axiom}[Independence of the Best Set, IBS] \label{IBS}
An opinion aggregator $f:\mathcal{O}(\mathfrak{X})\rightarrow  \mathcal{R}(X)$ satisfies the   \textit{IBS} axiom if, for every $x,y \in X$ and   $o^1, o^2 \in \mathcal{O}( \mathfrak{X} )$ such that the quotient orders are $M_{o^1}:\Sigma_{1} \succ \cdots  \succ \Sigma_{l}$ and $M_{o^2}:\Gamma_{1} \succ \cdots \succ \Gamma_{l'} \succ \Sigma_{2} \succ \cdots \succ \Sigma_{l}$ with $\Sigma_1=\Gamma_{1} \cup \cdots  \cup \Gamma_{l'}$, it holds that
\[
 \left\lbrack (x,y) \in P ( f(o^1)) 
  \Rightarrow (x,y) \in P \left( f\left( o^2 \right) \right) \right\rbrack.
\]
\end{axiom}
The IBS axiom is similar to the IWS axiom, but instead of focusing on the worst equivalence class, it concerns the independence of strict rankings from the partition of the best equivalence class.  
The IBS axiom is designed to prioritize opinions that exhibit low levels of support only within a limited number of subsets and is also a property examined in \cite{Bernardi2019} to penalize mediocrity in the context of social ranking.
It applies in situations where the best equivalence class initially contains multiple subsets, owing to indifference between the most preferred criteria. However, for various reasons (e.g., a change in the preference profile) ties among the most preferred criteria may be resolved, as illustrated in the following example.
\begin{example}\label{ex:ibs}
Consider $X, A, C, N$ as in Examples \ref{ex:2} and \ref{ex:3}; however, now $(\succsim'_i)_{i \in N}$ is such that
the Borda ranking based on the criterion $B_{\succsim'}$ is 
\[
B_{\succsim'}: d \sim' c \succ' b \succ' a \succ' f \succ' e.
\]
The state of opinion $o_{\succsim'}$ yields 
\[
\begin{array}{l}
m_{o_{\succsim'}}(\mathrm{Tr}(a))=7,\\
m_{o_{\succsim'}}(\mathrm{Tr}(b))=8,\\
m_{o_{\succsim'}}(\mathrm{Tr}(c))=9,\\
m_{o_{\succsim'}}(\mathrm{Tr}(d))=9,\\
m_{o_{\succsim'}}(\mathrm{Tr}(e))=5,\\
m_{o_{\succsim'}}(\mathrm{Tr}(f))=6,
\end{array}
\]
and $m_{o_{\succsim'}}(S)=0$, for each $S \in \mathfrak{X} \setminus \{\mathrm{Tr}(a), \ldots, \mathrm{Tr}(f)\}$. Therefore, the best equivalence class according to $m_{o_{\succsim'}}$ is 
\[
 \{\mathrm{Tr}(c), \mathrm{Tr}(d)\}.
\]
Now, consider $X, A, C, N$, $(\succsim_i)_{i \in N}$ and $o_{\succsim}$ as in Examples \ref{ex:2} and \ref{ex:3}. Note that the two best equivalence classes according to $m_{o_\succsim}$ are
\[
\{\mathrm{Tr}(d)\} \succ  \{\mathrm{Tr}(c)\}.
\]
Since $\{\{\mathrm{Tr}(d)\},  \{\mathrm{Tr}(c)\}\}$ constitutes a partition of $\{\{\mathrm{Tr}(d), \mathrm{Tr}(c)\}\}$, and all other subsets in $\mathfrak{X} \setminus \{\mathrm{Tr}(c) \mathrm{Tr}(d)\}$  are ranked identically by both $m_{o_{\succsim'}}$ and $m_{o_\succsim}$, the IBS axiom requires that if an opinion aggregator ranks Kemeny  strictly higher than Plurality in $o_{\succsim'}$ (which corresponds to $o^1$ according to the notations used in Axiom \ref{IBS}), then such an opinion aggregator must also rank Kemeny strictly higher than Plurality in $o_{\succsim}$ (which corresponds to $o^2$ according to the notations used in Axiom \ref{IBS}).

\end{example}

\begin{axiom}[Weak Intersection Very Important Player, WIVIP] \label{WIVIP}
An  opinion aggregator $f:\mathcal{O}(\mathfrak{X})\rightarrow  \mathcal{R}(X)$
satisfies the \textit{WIVIP} axiom if, for every $x,y \in X$ and
  $o \in \mathcal{O}( \mathfrak{X} )$ with respect to the q.o. of $M_{o}$ is
  \[
  M_{o}:\Sigma_{1} \succ \Sigma_{2},
  \]
with  $x \in \bigcap  \Sigma_{1}$ and 
  $y \in \left( \bigcap \Sigma_{1} \right)^{c}$, it holds that
\[
(x,y) \in P\left( f(o) \right).
\]
\end{axiom}
The WIVIP axiom addresses situations where the quotient order associated with a state of opinion $o$ contains only two classes and states that an element $x$ belonging to all the subsets in the best equivalence class should be ranked strictly  higher than any other element $y$ that does not exhibit the same behavior as $x$. Thus, elements 
 $x \in \bigcap  \Sigma_{1}$ is a {\it veto} element for the subsets in $\Sigma_{1}$ because its removal worsens the position of any subset in $\Sigma_{1}$, 
 and the WIVIP property states that veto elements should be ranked strictly higher than non-veto elements. 

A particular class of quotient orders with only two equivalence classes is a {\it simple game }, where $\Sigma_1$ and $\Sigma_2$ are such that the entire set of elements $X$ belongs to $\Sigma_1$ and if any subset $S$ belongs to $\Sigma_1$, then every   $T$ superset of $S$ also belongs to $\Sigma_1$.  Well-known solutions for Transferable Utility (TU) games (e.g., the Shapley value \cite{Shapley1953},  Banzhaf value \cite{BanzhafIII1964}, etc. See, for instance, \cite{owen95} for a general introduction) allocate at least as much utility to veto elements in simple games as any other non-veto element. Therefore, the WIVIP can be interpreted as a request to apply the principle of greater relevance to veto elements in simple games to any dichotomous quotient order.

\begin{example}\label{ex:wivip}
Consider $X, A, C$ as defined in Examples \ref{ex:2} and \ref{ex:3}, along with a state of opinion $o_{\succsim'}$ on $(\succsim'_i)_{i\in N}$ that provides the support
\[
\begin{array}{l}
m_{o_{\succsim'}}(\mathrm{Tr}(a))=
m_{o_{\succsim'}}(\mathrm{Tr}(b))=
m_{o_{\succsim'}}(\mathrm{Tr}(c))=
m_{o_{\succsim'}}(\mathrm{Tr}(d))=
m_{o_{\succsim'}}(\mathrm{Tr}(e))>0,
\end{array}
\]
and $m_{o_{\succsim'}}(S)=0$, for each $S \in \mathfrak{X} \setminus \{\mathrm{Tr}(a), \ldots, \mathrm{Tr}(e)\}$. Therefore, the best equivalence class according to $m_{o_{\succsim'}}$ is 
\[
\Sigma_1= \{\mathrm{Tr}(a), \ldots, \mathrm{Tr}(e)\},
\]
and the veto elements are both Copeland and Kemeny, which are the only two voting rules in Table \ref{tab:comparison} that verify all criteria $a, \ldots, e$.
An aggregation function satisfying the WIVIP axiom should rank these two veto elements strictly higher than any other element in $X$.
\end{example}

\begin{axiom}[Independence of Non-Unanimous Improvement, INUI] \label{INUI}
An opinion aggregator $f:\mathcal{O}(\mathfrak{X})\rightarrow  \mathcal{R}(X)$
satisfies the   \textit{INUI} axiom if, for every $x,y \in X$ and
  $o^1, o^2 \in \mathcal{O}( \mathfrak{X} )$ such that the q.o. of $M_{o^1}$ is 
  \[
  M_{o^1}:\Sigma_{1} \succ \cdots \succ \Sigma_{k} \succ \cdots \succ 
  \Sigma_{l}
  \]
  and 
the q.o. of $M_{o^2}$ is 
\[
M_{o^2}:\Sigma_{1} \succ \cdots \succ \Sigma_{k - 1} \succ \Delta \succ 
\Sigma_{k} \setminus \Delta \succ \Sigma_{k + 1}\cdots \succ \Sigma_{l},
\]
where $\Delta \subseteq \Sigma_{k}$ is such that
 $x \notin \bigcap \Delta$ and  $y \notin \bigcap \Delta$, 
  it holds that
\[
 \left\lbrack (x,y) \in f(o^1) 
  \Longleftrightarrow (x,y) \in  f( o^2) \right\rbrack.
\]
\end{axiom}
The INUI axiom states that the final ranking between two alternatives, $x$ and $y$, should remain unchanged when the support of subsets in a collection $\Delta$ (such that neither $x$ nor $y$ belongs to all the subsets in $\Delta$) is slightly improved. This is a strong condition requiring only a unanimous improvement in an alternative to influence its final ranking. This axiom is useful in scenarios where the objective is to reduce the impact of changes in voters' preferences on criteria that affect alternatives inconsistently, as illustrated in the following example:
\begin{example}\label{ex:INUI}
Consider $X, A, C$ as defined in Examples \ref{ex:2} and \ref{ex:3}, and a state of opinion $o_{\succsim'}$ on $(\succsim'_i)_{i\in N}$ that provides the support 

\begin{align*}
m_{o_{\succsim'}}(\mathrm{Tr}(d))>
m_{o_{\succsim'}}(\mathrm{Tr}(f))>
m_{o_{\succsim'}}(\mathrm{Tr}(a))&= m_{o_{\succsim'}}(\mathrm{Tr}(b))\\
&= m_{o_{\succsim'}}(\mathrm{Tr}(c))> m_{o_{\succsim'}}(\mathrm{Tr}(e))>0,
\end{align*}

\noindent  and $m_{o_{\succsim'}}(S)=0$, for each $S \in \mathfrak{X} \setminus \{\mathrm{Tr}(a), \ldots, \mathrm{Tr}(f)\}$.

Now, consider a change in the state of opinion such that (for instance, owing to a change in some voters' preferences)
we have a new state of opinion $o_{\succsim''}$ on $(\succsim''_i)_{i \in N}$, which yields support
\begin{align*}
m_{o_{\succsim''}}(\mathrm{Tr}(d))>
m_{o_{\succsim''}}(\mathrm{Tr}(f))>
m_{o_{\succsim''}}(\mathrm{Tr}(a))&=
m_{o_{\succsim''}}(\mathrm{Tr}(b))\\ &
>
m_{o_{\succsim''}}(\mathrm{Tr}(c))>
m_{o_{\succsim''}}(\mathrm{Tr}(e))>0,
\end{align*}

\noindent and $m_{o_{\succsim''}'}(S)=0$, for each $S \in \mathfrak{X} \setminus \{\mathrm{Tr}(a), \ldots, \mathrm{Tr}(f)\}$.
The INUI axiom requires that the slight improvement of the support of $\mathrm{Tr}(a)$ and $\mathrm{Tr}(b)$ from $o_{\succsim'}$ to $o_{\succsim''}$ should not affect the relative ranking between voting rules outside their intersection (such as Dodgson, Maximin, and Borda), as improvements in the support of these subsets $\mathrm{Tr}(a)$ and $\mathrm{Tr}(b)$ do not lead to a unanimous support of these voting rules.
\end{example}

\section{Characterization of the IIS opinion aggregator}
\label{sec3.2}

In this section, we introduce and discuss the axiomatic characterization of IIS opinion aggregator (for the axiomatic characterization of $f^{S}$, refer to \cite{Suzuki}).

First, we propose a series of lemmas showing the properties of
$e_{o}(x)$ in the relational axioms presented in Section \ref{sec:axioms}. These axioms are useful to prove the `if part' of Theorem \ref{theo3}, which is the main result of this section. 

\begin{lemma}[upper bound of $\mathbf{e}_{\mathbf{o}}(\mathbf{\cdot} )$]
\label{lem3}\normalfont%
For any $o \in \mathcal{O}\left( \mathfrak{X} \right)$ with
$M_{o}$: $\Sigma_{1} \succ \cdots \succ \Sigma_{l}$ and $x \in X$, 
we obtain $e_{o}(x) < l$.
\end{lemma}

\begin{proof of lemma}[\ref{lem3}]
\label{poflem3}\normalfont
Because $|X| \geq 3$, there exists 
$S \in \mathfrak{X} \setminus \mathfrak{X}\lbrack x\rbrack$. Moreover, because $\Sigma_{1},\cdots,\Sigma_{l}$ is a partition of $\mathfrak{X}$,
$\exists k \in \lbrack l\rbrack$ such that $S \in \Sigma_{k}$. 
If $S \in \Sigma_1$, then $e_{o}(x)=0<l$. Otherwise, from the definition of $e_{o}(x)$, it follows that $e_{o}(x) < k \leq l$. 
\end{proof of lemma}

For every $o \in \mathcal{O}\left( \mathfrak{X} \right)$
where $M_{o}$: $\Sigma_{1} \succ \cdots \succ \Sigma_{l}$ and for each $x \in X$, let $x_k$ be the number of subsets containing $x$ in $\Sigma_k$. That is,
\begin{equation}\label{def:xk}
x_k=|\{S\in \Sigma_k: x \in S\}|,
\end{equation}
for each value of $k \in \{1, \ldots, l\}$.

\begin{lemma}[property of $\boldsymbol{e_o}(\cdot)$ with respect to NT]
\label{lem5}
\normalfont%
For any $o \in \mathcal{O}\left( \mathfrak{X} \right)$, any permutation $\pi:X\to X$ and
$x \in X$, then  $e_{o}(x)= e_{o^\pi}\big(\pi(x)\big)$. Moreover, $\forall y\in X$, if $e_{o}(x) \geq e_{o}(y)$, then $e_{o^\pi}\big(\pi(x)\big) \geq e_{o^\pi}\big(\pi(y)\big)$.
\end{lemma}

\begin{proof of lemma}[\ref{lem5}]
\label{poflem5}\normalfont
The lemma follows immediately from the observation that 
\[
x\in\bigcap\Sigma_k \iff \pi(x)\in \bigcap\pi(\Sigma_k),\quad\forall k\in[l],
\]
where $\pi(\Sigma_k)=\{\pi(S): S\in\Sigma_k\}$.
\end{proof of lemma}

\begin{lemma}[property of $\mathbf{e}_{\mathbf{o}} (\mathbf{\cdot})$ with respect to IWS]
\label{lem8}\normalfont%
For any $o$, $o'\in\mathcal{O} (\mathfrak{X})$ such that $M_{o}$: 
$\Sigma_{1} \succ \cdots \succ \Sigma_{l - 1} \succ \Sigma_{l}$ 
and $M_{o'}:\Sigma_{1} \succ \cdots \succ \Sigma_{l - 1}\, 
\succ \Gamma_{1} \succ \cdots \succ \Gamma_{l'}$ 
and for any $x \in X$, we obtain:
\begin{enumerate}
\def\labelenumi{(\roman{enumi})}
\item
  if $e_{o}(x) = l - 1$, then $e_{o'}(x) \geq l - 1$,
\item
  if $e_{o}(x) < l - 1$, then $e_{o'}(x) = e_{o}(x)$.
\end{enumerate}
\end{lemma}
The proof of Lemma \ref{lem8} follows directly from the definition of
$e_{o}( \cdot )$ and is therefore omitted.

\begin{lemma}[property of $\boldsymbol{e_o}(\cdot)$ with respect to IBS]
\label{lem7}\normalfont%
For any $o$, $o'\in\mathcal{ O}(\mathfrak{X})$ such that $M_{o}$: 
$\Sigma_{1} \succ \cdots \succ \Sigma_{l}$ and $M_{o'}$: 
$\Gamma_{1} \succ \cdots \succ \Gamma_{l'} \succ \Sigma_{2} 
\succ \cdots \succ \Sigma_{l}$ and for any $x \in X$, we obtain:
\begin{enumerate}
\def\labelenumi{(\roman{enumi})}
\item
  if $e_{o}(x) > 0$, then $e_{o'}(x) = l' - 1 + e_{o}(x)$,
\item
  if $e_{o}(x) = 0$, then $e_{o'}(z) \leq l' - 1$.
\end{enumerate}
\end{lemma}

\begin{proof of lemma}[\ref{lem7}]
\label{poflem7}\normalfont
\mbox{}
\begin{enumerate}
\def\labelenumi{(\roman{enumi})}
\item
  Suppose $e_{o}(x) > 0$, then $x \in \bigcap\Sigma_{1} 
  = \bigcap ( \Gamma_{1} \cup \cdots \cup \Gamma_{l'} )$.
  Thus, $x \in \bigcap\Gamma_{k}$ for all $k = 1,\cdots,l'$.
  Furthermore, by definition of $e_{o}(x)$, we obtain
  $x \in \bigcap \Sigma_{k}$ for all $k \leq e_{o}(x)$ and
  $x \notin \bigcap \Sigma_{e_{o}(x) + 1}$. 
  Hence, it holds $e_{o'}(x) = l' - 1 + e_{o}(x)$.
\item
  Suppose $e_{o}(x) = 0$, then $x \notin \bigcap \Sigma_{1} = \bigcap 
  (\Gamma_{1} \cup \cdots \cup \Gamma_{l'} )$. 
  This implies that there exists $k \leq l'$ such that 
  $x \notin \bigcap \Gamma_{k}$. Hence, it holds $e_{o'}(x) < l'$.  
\end{enumerate}
\end{proof of lemma}

\begin{lemma}[property of $\mathbf{e}_{\mathbf{o}} (\mathbf{\cdot})$ with respect to WIVIP]
\label{lemwivip}\normalfont%
For any $o\in \mathcal{O} (\mathfrak{X})$ such that $M_{o}$: 
$\Sigma_{1} \succ \Sigma_{2}$ and $x,y \in X$ such that $x \in \bigcap  \Sigma_{1}$ and 
  $y \in \left( \bigcap \Sigma_{1} \right)^{c}$, we obtained $e_{o}(x)=1$ and $e_{o}(y)=0$.
\end{lemma}
The proof of Lemma \ref{lemwivip} follows directly from the definition of
$e_{o}( \cdot )$ and is thus omitted.

\begin{lemma}[property of $\boldsymbol{e_o}(\cdot)$ with respect to INUI]
\label{lam6}\normalfont%
For any $o,o' \in$\linebreak
$\mathcal{O}( \mathfrak{X} )$ with $M_{o}$: 
$\Sigma_{1} \succ \cdots \succ \Sigma_{\widehat{k}} \succ 
\cdots \succ \Sigma_{l}$, $\Delta \subseteq \Sigma_{\widehat{k}}$,
$x \notin \bigcap \Delta$, and 
$M_{o'}$: $\Sigma_{1} \succ \cdots \succ \Sigma_{\widehat{k} - 1} 
\succ \Delta \succ \Sigma_{\widehat{k}} \setminus \Delta 
\succ \Sigma_{\widehat{k} + 1} \succ \cdots \succ \Sigma_{l}$,
we obtain $e_{o}(x) = e_{o'}(x)$.
\end{lemma}

\begin{proof of lemma}[\ref{lam6}]
\label{poflem6}\normalfont
Note that $x \notin \bigcap \Delta$ and
$\Delta \subseteq \Sigma_{\widehat{k}}$ implies that
$x \notin \bigcap \Sigma_{\widehat{k}}$. Hence,
$e_{o}(x) < \widehat{k}$. If $e_{o}(x) < \widehat{k} - 1$, then for
each $k = 1,2,\cdots,e_{o}(x) + 1$, the $k^{\mathrm{th}}$ equivalence class
of $M_{o}$ must equal that of $M_{o'}$. Hence, it follows that
$e_{o'}(x) = e_{o}(x)$. If $e_{o}(x) = \widehat{k} - 1$, then
it must belong to $x \in \bigcap \Sigma_{k}$ for all
$k \leq \widehat{k} - 1$ and $x \notin \bigcap \Delta$ by
assumption. Therefore, $e_{o'}(x) = \widehat{k} - 1$. 
\end{proof of lemma}

This section introduces the main results of this study.
\begin{theorem}
\label{theo3}
\textit{Opinion aggregator} $f$ \textit{satisfies axioms NT,  IWS, IBS,
WIVIP and INUI if and only if} $f = f^{IIS}$.
\end{theorem}

The `if' part (i.e., $f^{IIS}$ satisfies the five axioms: NT, IWS, IBS,
WIVIP and INUI) follow directly from Lemmas \ref{lem5}, \ref{lem8}, \ref{lem7}, \ref{lemwivip}, and \ref{lam6}, respectively.


\vspace{.5\baselineskip}
\noindent
\textbf{Proof of the `only if' part of Theorem~\ref{theo3}.}
Suppose that $f$ satisfies all five axioms. We prove that for any
$o \in \mathcal{O}\left( \mathfrak{X} \right)$ and $x,y \in X$,
\begin{enumerate}
\def\labelenumi{(\roman{enumi})}
\item
  if $e_{o}(x) = e_{o}(y)$, then $(x,y) \in I\left( f(o) \right)$, and
\item
  if $e_{o}(x) > e_{o}(y)$, then $(x,y) \in P\left( f(o) \right)$.
\end{enumerate}

\vspace{.5\baselineskip}
\noindent
\textbf{Proof of (i).} 
Suppose $M_{o}:\Sigma_{1} \succ \cdots \succ \Sigma_{l}$ 
and $e_{o}(x) = e_{o}(y) = \widehat{k}$. 
Consider equivalence classes $\Sigma_{\widehat{k} + 1}, \Sigma_{\widehat{k} + 2}, \cdots, 
\Sigma_{l - 1}$ and define states of opinions $o^{( \widehat{k} + 2 )}, 
o^{( \widehat{k} + 3 )},\cdots,o^{(l)}\in \mathcal{ O}( \mathfrak{X} )$
as follows (the bold type is used only to emphasize their
differences)
\begin{align*}
&
M_{o^{( \hat{k} + 2 )}}\colon \Sigma_{1} \succ 
 \cdots \succ \Sigma_{\widehat{k}} \succ 
  \mathbf{\Sigma}_{\widehat{\boldsymbol{k}} 
   + \mathbf{1}} \cup \mathbf{\Sigma}_{\widehat{\boldsymbol{k}} 
    + \mathbf{2}} \succ \Sigma_{\widehat{k} + 3} 
    \succ \cdots \succ \Sigma_{l}, \\
&
M_{o^{( \hat{k} + 3 )}} \colon 
   \Sigma_{1} \succ \cdots \succ \Sigma_{\widehat{k}} \succ 
    \mathbf{\Sigma}_{\widehat{\boldsymbol{k}} + \mathbf{1}} 
      \cup  \mathbf{\Sigma}_{\widehat{\boldsymbol{k}} + \mathbf{2}} \cup \mathbf{\Sigma}_{\widehat{\boldsymbol{k}} + \mathbf{3}} \succ \cdots \succ \Sigma_{l}, \\
& \vdots \\
&
M_{o^{(l - 1)}}\colon \Sigma_{1} \succ \cdots \succ 
\Sigma_{\widehat{k}} \succ \mathbf{\Sigma}_{\widehat{\boldsymbol{k}} 
+ \mathbf{1}}\mathbf{\cup \cdots \cup}\mathbf{\Sigma}_{\boldsymbol{l} - \mathbf{1}} \succ \Sigma_{l},\\
&
M_{o^{(l)}}\colon \Sigma_{1} \succ \cdots \succ \Sigma_{\widehat{k}} \succ \mathbf{\Sigma}_{\widehat{\boldsymbol{k}} + \mathbf{1}} \cup \cdots \cup
 \mathbf{\Sigma}_{\boldsymbol{l}}.
\end{align*}

As $e_{o}(x) = e_{o}(y) = \widehat{k}$, any set in
$\Sigma_{1},\cdots,\Sigma_{\widehat{k}}$ contains both $x$ and $y$, 
This means that
$\Sigma_{1},\cdots,\Sigma_{\widehat{k}}\subseteq  \mathfrak{X}
\lbrack x\rbrack \cap \mathfrak{X}\lbrack y\rbrack$.

 Therefore, even if we swap the positions of $x$ and $y$ in $o^{(l)}$, the top $\hat{k}$ equivalence classes remain the same; accordingly, the remaining equivalence class, $\Sigma_{\hat{k}+1} \cup \cdots \cup \Sigma_l$, should also be the same. Therefore, NT implies that $(x,y)\in R(f(o^{(l)})) \Longleftrightarrow (y,x)\in R(f(o^{(l)}))$. Since $R(f(o^{(l)})) \in \mathcal{R}(X)$ is assumed to be complete, we can conclude that $(x,y)\in I(f(o^{(l)}))$.

By the definition, for each $k=1,2, \cdots, l-\hat{k}-2$, $M_{o^{(l-k)}}$ is obtained from $M_{o^{(l)}}$ by improving the position of $\Sigma_{\hat{k}+1} \cup \cdots \cup \Sigma_{l-k}$. Since $x, y \notin \bigcap \Sigma_{\hat{k}+1}$ by definition of $\hat{k}\left(=e_{\succsim}(x)=e_{\succsim}(y)\right)$, we have also that $x, y \notin \bigcap\left(\Sigma_{\hat{k}+1} \cup \cdots \cup \Sigma_{l-k}\right)$. Hence, INUI implies that $(x,y)\in I(f(o^{(l-k)}))\iff (x,y)\in I(f(o^{(l-k-1)}))$, for all $k=0,1, \cdots, l-\hat{k}-2$ with $o^{(\hat{k}+1)}=o$, and then $(x,y)\in I(f(o))$.

\vspace{.5\baselineskip}
\noindent
\textbf{Proof of (ii).} 
Suppose $M_{o}$: $\Sigma_{1} \succ \cdots \succ \Sigma_{l}$ and
$\widehat{k} = e_{o}(x) > e_{o}(y)$. We construct $o'$ by reducing
the support of the subsets in $\Sigma_{1},\cdots,\Sigma_{\widehat{k} - 1}$ while
increasing the support of the subsets in $\Sigma_{\widehat{k} + 2},\cdots,\Sigma_{l}$
ensuring that 
\[
M_{o'}: \Sigma_{1} \cup \cdots \cup \Sigma_{\widehat{k}} 
\succ \Sigma_{\widehat{k} + 1} \cup \cdots \cup \Sigma_{l}.
\]
As $\widehat{k} = e_{o}(x) > e_{o}(y)$, it follows that
$x \in \bigcap {\Sigma_{1} \cup \cdots \cup \Sigma_{\widehat{k}}}$
and $y \notin \bigcap  {\Sigma_{1} \cup \cdots \cup \Sigma_{\widehat{k}}}$.
Thus, WIVIP implies that 
$(x,y) \in P\left( f\left( o' \right) \right)$.

We define a new state of opinion 
$o''\in \mathcal{ O}\left( \mathfrak{X} \right)$ by increasing 
support of the subsets in $\Sigma_{1},\cdots,\Sigma_{\widehat{k} - 1}$ in a manner that
\[
M_{o^{''}}: \Sigma_{1} \succ \cdots \succ \Sigma_{\widehat{k}} \succ \Sigma_{\widehat{k} + 1} \cup \cdots \cup \Sigma_{l}.
\]
By applying IBS, where $o{'}$ takes the role of $o^1$ and $o{''}$ takes the role of $o^2$, it follows that $(x,y) \in P\left( f\left(o{''} \right) \right)$.

Similarly, using IWS, where $o{''}$ serve as $o^1$ and $o$ serve as $o^2$, we obtain $(x,y) \in P\left( f(o) \right)$. 
This completes the proof of (ii).\\ \\

Finally, we observe some similarities between the IIS opinion aggregator $f^{IIS}$ and the lex-cel social ranking introduced in \cite{Bernardi2019}. We recall that the lex-cel can be computed using a support relationship. For any $o \in \mathcal{O}( \mathfrak{X} )$ and the corresponding support relation $M_o:\Sigma_{1} \succ \cdots \succ \Sigma_{l}$, for each $x \in X$, we define the vector as follows:
\begin{equation}\label{def:lexcelo}
    \theta_o(x)=(x_1, \ldots, x_l),
\end{equation} 
where $x_k$ is as defined in \eqref{def:xk}. 
The lex-cel on $M_o$ is defined as the weak order on $R_{le}^o \subseteq X \times X$ such that
\[
(x,y) \in R_{le}^o \Longleftrightarrow \theta_o(x) \geq_L \theta_o(y),
\]
for all $x,y \in X$ and where $\geq_L$ denotes the standard lexicographic comparison between vectors (i.e., $\theta_{o}(x) \ge_L \theta_{o} (y)$ if either  $\theta_{o}(x)=\theta_{o}(y)$ or  there exists $t$ such that $x_t>y_t$ and  $x_r=y_r$  for all $r \in \{1,\dots, l-1\}$).

For instance, if we consider $M_{o_\succsim}$ again in Example \ref{ex:3}, we obtain 
\[\theta_{o_\succsim}(Approval)=(1,0,0,0,1,0,62), \text{ and } \theta_{o_\succsim}(Borda)=(1,0,1,0,1,1,60),
\] and thus, $(Borda, Approval) \in P(R_{le}^{o_\succsim})$. For comparison with IIS, we provide the ranking $R_{le}^{o_\succsim}$ obtained with lex-cel:
\begin{multline*}
    R_{le}^{o_\succsim}:
Copeland, Kemeny \succ Maximin \succ Plurality\\ \succ Borda \succ Approval \succ Dodgson.
\end{multline*}

The IIS function follows a lexicographic criterion that prioritizes the superiority of alternatives in $X$. It is easy to verify that a strict preference according to IIS implies a strict relation according to lex-cel; that is, if $(x,y) \in P(f^{IIS}(o))$, then $(x,y) \in P(R_{le}^o)$. Moreover, if we obtain the equivalence according to lex-cel $(x,y) \in I(R_{le}^o)$, which is equivalent to stating that $\theta_o(x)=\theta_o(y)$, then equivalence also holds according to IIS $(x,y) \in P(f^{IIS}(o))$.

The function’s tendency to reward the excellence of alternatives can also be inferred from the analysis of the axioms satisfied by both the IIS and lex-cel. As noted in \cite{Bernardi2019}, it is easy to verify that lex-cel also satisfies the IWS, WIVIP, and NT axioms, which may be interpreted as properties aimed at prioritizing  alternatives belonging to subsets with a large support. However, unlike the lex-cel, $f^{IIS}$ also satisfies the IBS axiom, which aims to penalize alternatives belonging to multiple subsets with low support (see \cite{Bernardi2019} for an extended discussion of IBS and its role in characterizing the dual lex-cel). In this respect, the IIS aggregation function can be considered a variation of lex-cel, designed to further reward alternatives that belong to subsets of criteria with high support while avoiding undue relevance to their position in subsets with moderate support.

\subsection{Independence of the axioms}
\label{sec3.3}

We examine the logical independence of the axioms in Theorem~\ref{theo3}, as proven 
in Proposition \ref{prop1}. To this end, we define new opinion aggregators.

Let $\vartriangleright$ be the linear order of $X$. We define an
opinion aggregator called the \textit{IIS rule with
tie breaking using } $\vartriangleright$, denoted by
$f^{IIS, \vartriangleright}$, as follows: for any
$o \in \mathcal{O}( \mathfrak{X})$ and $x,y \in X$, 
we obtain $(x,y) \in f^{IIS, \vartriangleright}(o)$ if and only if one of the
the followings hold:
\begin{enumerate}
\def\labelenumi{(\roman{enumi})}
\item
  $e_{o}(x) > e_{o}(y)$,
\item
  $e_{o}(x) = e_{o}(y) \in \{ 0,l - 1\}$,
\item
  $0 < e_{o}(x) = e_{o}(y) < l - 1$ and $x \vartriangleright y$.
\end{enumerate}
For $o \in \mathcal{O}(\mathfrak{X})$ 
where $M_{o}$: $\Sigma_{1} \succ \cdots \succ \Sigma_{l}$, 
take $\theta_{o}$ as in \eqref{def:lexcelo}. Let
$\tau_{o}(x) := \left( {\dot{x}}_{1},\cdots,{\dot{x}}_{l} \right)$,
where ${\dot{x}}_{k} := \sum_{i \leq k} x_{i}$ and the elements $x_i$ are defined in \eqref{def:xk}. For two
vectors $a := \left( a_{1},a_{2},\cdots,a_{p} \right), 
b := \left( b_{1},b_{2},\cdots, b_{p} \right) \in \mathbb{Z}^{p}$,
recall that  $a \geq^{L}b$ denotes the standard lexicographic comparison
between the vectors $a$ and $b$
Thus, we define $a >^{L}b$ if $a \geq^{L}b$ and
$\neg\left( b \geq^{L}a \right)$.

We define an opinion aggregator, referred to as the \textit{IIS rule with tie breaking using } $\tau$, denoted as
$f^{IIS,\tau}$, as follows: for any
$o \in \mathcal{O}\left( \mathfrak{X} \right)$ and $x,y \in X$, 
we obtain $(x,y) \in f^{IIS,\tau}$ if and only if one of the following holds:
\begin{enumerate}
\def\labelenumi{(\roman{enumi})}
\item
  $e_{o}(x) > e_{o}(y)$,
\item
  $e_{o}(x) = e_{o}(y) = 0$,
\item
  $e_{o}(x) = e_{o}(y) > 0$ and $\tau_{o}(x) \geq^{L}\tau_{o}(y)$.
\end{enumerate}
We define the opinion aggregator $f_{1}$, for any
$o \in \mathcal{O}\left( \mathfrak{X} \right)$, by
\[
f_{1}(o)\colon 
 \left\{ x \in X\colon e_{o}(x) \geq 2 \right\} 
  \succ \left\{ x \in X\colon e_{o}(x) = 1 \right\} 
   \succ \left\{ x \in X\colon  e_{o}(x) = 0 \right\}.
\]
Let us define the opinion aggregator $f_{2}$, for any
$o \in \mathcal{O}\left( \mathfrak{X} \right)$ with
$M_{o}$: $\Sigma_{1} \succ \cdots \succ \Sigma_{l}$, by
\[  
f_{2}(o)\colon 
 \left\{ x \in X \colon e_{o}(x) = l - 1 \right\} 
  \succ \left\{ x \in X\colon e_{o}(x) < l - 1 \right\}.
\]
We define an opinion aggregator, referred to as the \textit{indifference rule}, 
denoted as $f^{I}$. For any $o \in \mathcal{O}\left( \mathfrak{X} \right)$, it is defined as 
$f^{I}(o) = X \times X$.

\begin{proposition}
\label{prop1}\normalfont%
The five axioms in Theorem~\ref{theo3} (NT, INUI, IWS, IBS, and 
and WIVIP) are logically independent. That is,  $f^{IIS, \vartriangleright}$
satisfies the five axioms except NT, $f^{IIS,\tau}$ satisfies the five
except INUI, $f_{1}$ satisfies the five except IBS, $f_{2}$
satisfies the five except IWS, and $f^{I}$ satisfies the five except WIVIP.
\end{proposition}

\begin{proof of proposition}[\ref{prop1}]\normalfont

\noindent\vspace{.5\baselineskip}\\
\textbf{On WIVIP:} 

The statement (all
$f^{IIS},f^{IIS,\tau},f^{IIS, \vartriangleright},f_{1},f_{2}$ satisfy
WIVIP but $f^{I}$ does not) is straightforward, based on the definition of
opinion aggregator. Furthermore, $f^{I}$ satisfies
other axioms. Hence, we proceed to prove the remaining part

\vspace{.5\baselineskip}
\noindent
\textbf{On NT:}

Let $\pi:X\to X$ be a permutation. From Lemma~\ref{lem5}, we obtain
$e_{o}(x) \geq e_{o}(y) \iff e_{o^\pi}(\pi(x)) \geq e_{o^\pi}(\pi(y))$. Thus, if $(x,y) \in f^{IIS}(o) \cap f_{1}(o) \cap f_{2}(o)$ iff $(\pi(x),\pi(y)) \in f^{IIS}(o^\pi) \cap f_{1}(o^\pi) \cap f_{2}(o^\pi)$. 
Furthermore, from Lemma~\ref{lem5}, 
we obtain that $\tau_{o}(x) \geq^{L}\tau_{o}(y)\iff \tau_{o^\pi}(\pi(x)) \geq^{L}\tau_{o^\pi}(\pi(y))$. Therefore, all $f^{IIS},f_{1},f_{2},$ and $f^{IIS,\tau}$ satisfy NT.

However, $f^{IIS, \vartriangleright}$ does not satisfy NT. A
counterexample is as follows. Suppose $M_{o}:\{ \{ x,y \} \} \succ \{ \{ x,y,z \} \} \succ \{ \{ x,y \},
\{ x,y,z \} \}^{c}$ and $x \vartriangleright y$. Let $\pi$ be a permutation that switches only $x$ and $y$, then 
$M_{o^\pi}=M_{o}$. In this case $(x,y) \in P\left( f^{IIS, \vartriangleright}(o) \right)$,  but $(\pi(x),\pi(y))=(y,x) \notin P\left( f^{IIS, \vartriangleright}(o) \right)$.

\vspace{.5\baselineskip}
\noindent
\textbf{On INUI:}

Take $o,o'\in\mathcal{O}(\mathfrak{X})$ with associated q.o. $M_{o}:\Sigma_{1} \succ \cdots \succ \Sigma_{\widehat{k} - 1} 
\succ \Sigma_{\widehat{k}} \succ \cdots \succ \Sigma_{l}$ and $M_{o'}:\Sigma_{1} \succ \cdots \succ \Sigma_{\widehat{k} - 1} \succ, 
\Delta \succ \Sigma_{\widehat{k}} \setminus \Delta 
\succ \cdots \succ \Sigma_{l}$, where $\Delta \subseteq \Sigma_{\hat{k}}$ with $x \notin \bigcap \Delta$ 
and $y \notin \bigcap \Delta$. 

From Lemma \ref{lam6}, we obtain
\begin{align}
(e_o(x), e_o(y))=(e_{o'}(x), e_{o'}(y)).
\label{eq8}
\end{align}
Thus, $f^{IIS}(o) | \left\{ x,y \right\} = f^{IIS}\left( o' \right) 
| \left\{ x,y \right\}$. 
This means that $f^{IIS}$ satisfies INUI. From (\ref{eq8}), 
$f^{IIS, \vartriangleright},f_{1},$ and $f_{2}$ also satisfy INUI; for 
$f^{IIS, \vartriangleright}$; note that $\vartriangleright$ does not
depend on $o$ and $o'$. Thus, 
$f^{IIS, \vartriangleright}(o) | \left\{ x,y \right\} 
= f^{IIS, \vartriangleright}\left( o' \right) | \left\{ x,y \right\}$ 
holds when we use (\ref{eq8}).

However, $f^{IIS,\tau}$ does not satisfy INUI. A counterexample is as
follows. Let $x,y,z \in X$ and $\Sigma_{1} := \{ \{ x \}, 
\{ x,z \},\{ y \},\{ y,z \} \}$. 
Let $M_{o}$ be such that $M_{o}:\Sigma_{1} \succ \Sigma_{1}^{c}$. For 
$\Delta := \{ \{ x,z \},\{ y \},\{ y,z \} 
\}$, let $M_{o'}$: $\Delta \succ \Sigma_{1} \setminus 
\Delta \succ \Sigma_{1}^{c}$.
Subsequently, we obtain $(x,y) \in I( f^{IIS,\tau}(o) )$ and
$(y,x) \in P( f^{IIS,\tau}( o' ) )$. This contradicts INUI.

\vspace{.5\baselineskip}
\noindent
\textbf{On IBS:}\\
Take $o,o'\in\mathcal{O}(\mathfrak{X})$ with associated q.o. $M_{o}:\Sigma_{1} \succ \cdots \succ \Sigma_{l}$  
and $M_{o'}:\Gamma_{1} \succ \cdots \succ \Gamma_{l'} 
\succ \Sigma_{2} \succ \cdots \succ \Sigma_{l}$, respectively. Suppose $(x,y) \in P\left( f^{IIS}(o) \right)$. Subsequently,
$e_{o}(x) > e_{o}(y) \geq 0$. Hence, from Lemma \ref{lem7}, we can state that
$e_{o'}(x) > e_{o'}(y)$. This implies that $f^{IIS}$ satisfies
IBS.

We assume that $(x,y) \in P\left( f^{IIS,\tau}(o) \right)$. Subsequently, we obtain
either $e_{o}(x) > e_{o}(y)$ or [$e_{o}(x) = e_{o}(y) > 0$ and
$\tau_{o}(x) >^{L}\tau_{o}(y)$]. In the former case, we obtain
$e_{o'}(x) > e_{o'}(y)$ from Lemma \ref{lem7}. In the latter case, we obtain
$e_{o'}(x) = e_{o'}(y)$ from Lemma \ref{lem7}. Furthermore, since
$x,y \in \bigcap \Sigma_{1}$ (because
$e_{o}(x) = e_{o}(y) > 0$), we obtain
$\left| \Gamma_{k}\lbrack x\rbrack \right| = \left| \Gamma_{k}\lbrack y\rbrack \right| = \left| \Gamma_{k} \right|$
for all $k \in \left\lbrack l' \right\rbrack$. Hence,
$\tau_{o}(x) >^{L}\tau_{o}(y)$ implies that
$\tau_{o'}(x) >^{L}\tau_{o'}(y)$. To conclude, we obtain 
$(x,y) \in P\left( f^{IIS,\tau}\left( o' \right) \right)$ and then $f^{IIS,\tau}$ satisfies IBS.

Suppose that
$(x,y) \in P\left( f^{IIS, \vartriangleright}(o) \right)$. Then, we
either $e_{o}(x) > e_{o}(y)$ or
[$0 < e_{o}(x) = e_{o}(y) < l - 1$ and $x \vartriangleright y$].
From Lemma \ref{lem7}, it is straightforward to verify that the former implies that
$e_{o'}(x) > e_{o'}(y)$. The latter implies that
$0 < e_{o'}(x) = e_{o'}(y) < l' + l - 2$ and
$x \vartriangleright y$. Thus, we  also state that
$(x,y) \in P\left( f^{IIS, \vartriangleright}\left( o' \right) \right)$,
which means that $f^{IIS, \vartriangleright}$ also satisfies IBS.

We assume that $(x,y) \in P\left( f_{2}(o) \right)$. Subsequently, it follows that
$e_{o}(x) = l - 1 > e_{o}(y) \geq 0$. From Lemma \ref{lem7}, we obtain
$e_{o'}(x) = l' + l - 2$ and
$e_{o'}(y) \leq \max\left\{ l' - 1 + e_{o}(y),l' - 1 \right\} < l' + l - 2$.
Thus, $(x,y) \in P\left( f_{2}\left( o' \right) \right)$ and then $f_{2}$ satisfies IBS.

However, $f_{1}$ does not satisfy IBS. A counterexample is as follows. Let
$\Sigma_{1} := \left\{ \left\{ x,y \right\},\left\{ x,y,z \right\},\left\{ x \right\} \right\}$
and $M_{o}$ is $\Sigma_{1} \succ \Sigma_{1}^{c}$. Let
$M_{o'}$: $\left\{ \left\{ x,y \right\} \right\} \succ \left\{ \left\{ x,y,z \right\} \right\} \succ \left\{ \left\{ x \right\} \right\} \succ \Sigma_{1}^{c}$.
Now, we have $(x,y) \in P\left( f_{1}(o) \right)$ and
$(x,y) \in I\left( f_{1}(o) \right)$ (because $e_{o'}(x) \geq 2$
and $e_{o'}(y) \geq 2$). This contradicts IBS.

\vspace{.5\baselineskip}
\noindent
\textbf{On IWS:}\\
Take $o,o'\in\mathcal{O}(\mathfrak{X})$ with associated q.o. $M_{o}:\Sigma_{1} \succ \cdots \succ \Sigma_{l - 1} \succ \Sigma_{l}$and $M_{o'}:\Sigma_{1} \succ \cdots \succ \Sigma_{l - 1} \succ \Gamma_{1} \succ \cdots \succ \Gamma_{l'}$, respectively. Suppose $(x,y) \in P\left( f^{IIS}(o) \right)$. Then,
$(l - 1 \geq )e_{o}(x) > e_{o}(y)$. Hence, from Lemma~\ref{lem8},
$e_{o'}(x) > e_{o'}(y)$. 
Therefore, $(x,y) \in P\left( f^{IIS}\left( o' \right) \right)$. 
Thus, $f^{IIS}$ satisfies IWS.

We assume that $(x,y) \in P\left( f^{IIS,\tau}(o) \right)$. Subsequently, we obtain
either $e_{o}(x) > e_{o}(y)$ or [$e_{o}(x) = e_{o}(y) > 0$ and
$\tau_{o}(x) >^{L}\tau_{o}(y)$]. In the former case, Lemma \ref{lem8} implies that
$e_{o'}(x) > e_{o'}(y)$. Hence,
$(x,y) \in P\left( f^{IIS,\tau}\left( o' \right) \right)$. Consider
the latter case. Since
${\dot{x}}_{l} = 2^{|X| - 1} - 1 = {\dot{y}}_{l}$ by definition
$\tau_{o}(x) >^{L}\tau_{o}(y)$ implies that $\exists k < l$ such that
${\dot{x}}_{k} \neq {\dot{y}}_{k}$. Hence, it follows
that $\exists\widehat{k} < l$ such that $x_{\widehat{k}} \neq y_{\widehat{k}}$. 
At this $\widehat{k}$, we obtain $\min\left\{ x_{\widehat{k}}, 
y_{\widehat{k}} \right\} < \left| \Sigma_{\widehat{k}} \right|$.
This implies that either $e_{o}(x) < \widehat{k}$ or
$e_{o}(y) < \widehat{k}$. As $e_{o}(x) = e_{o}(y)$, it follows that
$e_{o}(x) = e_{o}(y) < \widehat{k} < l$. Therefore, from Lemma~\ref{lem8},
we can conclude that $0 < e_{o'}(x) = e_{o}(x) = e_{o}(y) = e_{o'}(y) \leq l - 2$.
The first $\widehat{k}$ equivalence classes are the same for
$M_{o}$ and $M_{o'}$, we obtain 
$\tau_{o'}(x) >^{L}\tau_{o'}(y)$. Hence,
$(x,y) \in P\left( f^{IIS,\tau}\left( o' \right) \right)$ and then $f^{IIS,\tau}$ satisfies IWS.

We assume that $(x,y) \in P\left( f^{IIS, \vartriangleright} \right)$. Then,
either $e_{o}(x) > e_{o}(y)$ or
[$0 < e_{o}(x) = e_{o}(y) < l - 1$ and $x \vartriangleright y$] is obtained.
From Lemma \ref{lem8}, it follows that $e_{o'}(x) > e_{o'}(y)$ or
[$0 < e_{o'}(x) = e_{o'}(y) < l - 1$ and
$x \vartriangleright y$]. Thus,
$(x,y) \in P\left( f^{IIS, \vartriangleright}\left( o' \right) \right)$ and then $f^{IIS, \vartriangleright}$ satisfies IWS.

We assume that $(x,y) \in P\left( f_{1}(o) \right)$. Subsequently, we obtain
$e_{o}(x) > e_{o}(y)$ and $e_{o}(y) \leq 1$. From Lemma \ref{lem3}, we obtain 
$l - 1 \geq e_{o}(x) > e_{o}(y)$. Hence, from Lemma \ref{lem8}, 
$e_{o'}(x) \geq e_{o}(x) > e_{o}(y) = e_{o'}(y) \leq 1$ is obtained.
Thus, $(x,y) \in P\left( f_{1}\left( o' \right) \right)$.

However, $f_{2}$ does not satisfy IWS. A counterexample is as follows. Let
$M_{o}$: $\left\{ \left\{ x \right\} \right\} \succ \left\{ \left\{ x \right\} \right\}^{c}$
and
$M_{o'}$: $\left\{ \left\{ x \right\} \right\} \succ \left\{ \left\{ y \right\} \right\} \succ \left\{ \left\{ x \right\},\left\{ y \right\} \right\}^{c}$.
Subsequently, we obtain $(x,y) \in P\left( f_{2}(o) \right)$ and
$(x,y) \in I\left( f_{2}\left( o' \right) \right)$.

\end{proof of proposition}

\section{Relationship between CSCCs and opinion aggregators}
\label{sec3}

\begin{figure}[t]
\begin{center}
 \scalebox{1}{\includegraphics[bb=0 0 197 133]{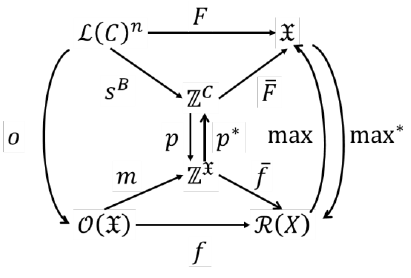}}%
\end{center} 
\caption{Relationship of the functions}
\label{f1}
\end{figure}

To examine the relationship between CSCC and opinions
aggregators, we present a theoretical view of these notions (see
Figure \ref{f1}).
\begin{itemize}
\item[-]
  Let $s^{B}$: $\mathcal{L}(C)^{n} \rightarrow \mathbb{Z}^{C}$ as 
  $s^{B}( \succsim )(c) := s_{\succsim}^{B}(c)$ for all $c \in C$. 
  By definition of $s_{\succsim}^{B}$, we obtain $s^{B}( \succsim )(c) 
  := \sum_{i \in N} | \{ d \in C\colon c \succsim_{i}d \} | 
  = \sum_{i \in N} | \{ T \in \mathfrak{X}\colon c \succsim_{i}
  \mathrm{Tr}^{*}(T) \} |$.
\item[-]
  Let $o$: $\mathcal{L}(C)^{n}\rightarrow \mathcal{O}( \mathfrak{X} )$
  as $o( \succsim) := o_{\succsim}$ for all 
  $\succsim \in \mathcal{ L}(C)^{n}$. By definition of $o_{\succsim}$, 
  we have $o( \succsim )(S,T) := | \{ i \in N\colon  \mathrm{Tr}^{*}(S) 
  \succsim_{i} \mathrm{Tr}^{*}(T) \} |$.
\item[-]
  Let
  $m$: $\mathcal{O}(\mathfrak{X} ) \rightarrow \mathbb{Z}^{\mathfrak{X}}$
  as $m(o)(S) := m_{o}(S)$ for all
  $o \in \mathcal{O}( \mathfrak{X} )$ and $S \in \mathfrak{X}$.
\item[-]
  Let $\max$: $\mathcal{R}(X) \rightarrow  \mathfrak{X}$ as
  $\max(R) := \{ x \in X\colon (x,y) \in R,\ \forall y \in X \}$
  for all $R\in \mathcal{R}(X)$.
\item[-]
  Let $\max^{*}$: $\mathfrak{X} \rightarrow  \mathcal{R}(X)$ as
  $\max^{*}(S)\colon S \succ S^{c}$, for all $S \in \mathfrak{X}$.
\item[-]
  Let $p$: $\mathbb{Z}^{C} \rightarrow \mathbb{Z}^{\mathfrak{X}}$ as, for
  all $g \in \mathbb{Z}^{C}$ and $S \in \mathfrak{X}$, we have
\end{itemize}
\begin{align*}
p(g)(S) 
 := \left\{ 
 \begin{array}{@{}ll@{}}
  g( \mathrm{Tr}^{*}(S) ) 
  & \mathrm{if}\ \mathrm{Tr}^{*}(S) \in C, \\
  0&  \mathrm{otherwise.}
\end{array}
\right.
\end{align*}
\begin{itemize}
\item[-]
  Let $p^{*}$: $\mathbb{Z}^{\mathfrak{X}} \rightarrow \mathbb{Z}^{C}$ be
  such that, for all $h \in \mathbb{Z}^{\mathfrak{X}}$ and $c \in C$,
  we obtain
\[  
p^{*}(h)(c) := h( \mathrm{Tr}(c) ).
\]
\end{itemize}
By definition, we obtain 
\begin{align}
&\max \circ\, {\max} ^*=1_\mathfrak{X};\label{eq2} \\
& p^* \circ p=1_{\mathbb{Z}} c.  \label{eq3}
\end{align}
Identity \eqref{eq2} is straightforward. For all 
$g \in \mathbb{Z}^{C}$ and $c \in C$, we obtain
\begin{align*}
(p^{*} \circ p)(g)(c)
 = p^{*}( p(g) )(c) = p(g)( \mathrm{Tr}(c) ) 
  = g( \mathrm{Tr}^{*}( \mathrm{Tr}(c) ) ) = g(c).
\end{align*}
This indicates that identity \eqref{eq3} holds.

From Proposition \ref{prop:bordamo} we obtain
\begin{align}
m \circ o=p \circ s^B .
\label{eq4}
\end{align}
In the following definition, we introduce classes of CSCCs and opinion aggregators.
\begin{definition}
\label{def7}\normalfont%
\mbox{}
\begin{enumerate}
\def\labelenumi{(\roman{enumi})}
\item
  CSCC $F$ is \textit{Borda-based} if 
  $\exists\, \overline{F}$: $\mathbb{Z}^{C}\rightarrow \mathfrak{X}$ 
  such that $F = \overline{F} \circ s^{B}$.
\item
  Opinion aggregator $f$ satisfies \textit{anonymity} if
  $\exists\, \overline{f}$: $\mathbb{Z}^{\mathfrak{X}} \rightarrow 
  \mathcal{R}(X)$ such that $f = \overline{f} \circ m$.
\end{enumerate}
\end{definition}

We define the relationship between CSCCs and opinion aggregators.
\begin{definition}
    CSCC $F$ is \textit{induced by an opinion aggregator}  $f$: 
  $\mathcal{O}\left( \mathfrak{X} \right) \rightarrow \mathcal{R}(X)$
  if $F = \max \circ f \circ o$,
\end{definition}
\begin{theorem}
\label{theo1}
\textit{A CSCC} $F$ \textit{is Borda-based if and only if it is induced by
anonymous opinion aggregator.}
\end{theorem}

\begin{proof of theorem}[\ref{theo1}]
\label{poftheo1}\normalfont%
\emph{`Only if' part}. If $F$ is Borda-based, then obtain
$F = \overline{F} \circ s^{B}$. Let $\overline{f} := \max^{*}  
\circ \overline{F} \circ p^{*}$. From (2), (3), and (4), we obtain
$\max \circ \overline{f} \circ m \circ o = \max \circ \left( \max^{*} 
\circ \overline{F} \circ p^{*} \right) \circ m \circ o 
= \overline{F} \circ p^{*} \circ m \circ o = \overline{F} \circ p^{*} 
\circ p \circ s^{B} = \overline{F} \circ s^{B} = F$.
This implies that $F$ is induced by an anonymous opinion aggregator.

\emph{`If' part}. If $F$ is induced by an anonymous opinion aggregator, then 
we obtain $F = \max \circ \overline{f} \circ m \circ o$. Let
$\overline{F} := \max \circ \overline{f} \circ p$. From (4), we obtain:
$\overline{f} \circ s^{B} = \max \circ \overline{f} \circ p \circ s^{B} 
= \max \circ \overline{f} \circ m \circ o = F$.
This implies that $F$ is Borda-based.
\end{proof of theorem}

Next lemma  is a straightforward consequence of 
equation~(\ref{eq4}) (Thus, the proof is omitted). 
This expresses the relationship between 
$m_{o_{\succsim}}( \cdot )$ and $s_{\succsim}^{B}( \cdot )$.

\begin{lemma}
\label{lem1}\normalfont%
The following holds:
\begin{enumerate}
\def\labelenumi{(\roman{enumi})}
\item
  For any $c \in C$ and $o \in \mathcal{O}\left( \mathfrak{X} \right)$, 
  then $m_{o_{\succsim}}\left( \mathrm{Tr}(c) \right) 
  = s_{\succsim}^{B}(c) 
  > 0$.
\item
  For any $S \in \mathfrak{X}  \setminus \mathrm{Tr}(C)$ 
  and $o \in \mathcal{O}\left( \mathfrak{X} \right)$, 
  then $m_{o_{\succsim}}(S) = 0$.
\end{enumerate}
\end{lemma}

\begin{lemma}
\label{lem2}\normalfont%
For any $\succsim \in \mathcal{ L}( C )^{n}$ 
where $B_\succsim:\Sigma_{1} \succ \cdots \succ \Sigma_{l}$, 
we obtain $M_{o_{\succsim}}$: $\Sigma_{1}' \succ \cdots \succ, 
\Sigma_{l + 1}'$, where
\begin{align}
\left\{
 \begin{array}{@{}ll@{}}
  \Sigma_k^{\prime}
   =\left\{\mathrm{Tr}(c)\colon c \in \Sigma_k\right\} & (1 \leq k \leq l), \\
   \Sigma_{l+1}^{\prime}
    =\mathfrak{X} \backslash \mathrm{Tr}(C).
\end{array}\right.
\label{eq5}
\end{align}
Furthermore, we obtain $T_{k} = T_{k}'$ for all $k = 1,\cdots,l$.
\end{lemma}

\begin{proof of lemma}[\ref{lem2}]
\label{poflem2}\normalfont
From Lemma~\ref{lem1}, for all $S \in \mathrm{Tr}(C)$ 
and $T \in \mathfrak{X } \setminus \mathrm{Tr}(C)$, 
we obatin $m_{o_{\succsim}}(S) > m_{o_{\succsim}}(T) = 0$. 
Hence, the worst equivalence class of $M_{o_{\succsim}}$ is $\mathfrak{X} \setminus \mathrm{Tr}(C)$.

From Lemma~\ref{lem1}, for all $c,d \in C$, we obtain 
$m_{o_{\succsim}}( \mathrm{Tr}(c) ) \geq m_{o_{\succsim}}(\mathrm{Tr}(d) ) \Leftrightarrow s_{\succsim}^{B}(c) 
\geq s_{\succsim}^{B}(d)$. 
This result implies that $\Sigma_{k}' = \left\{ \mathrm{Tr}(c)\colon 
c \in \Sigma_{k} \right\}$, for all 
$k = 1,\cdots,l$. Based on this, we obtain
$\Sigma_{l + 1}'\mathfrak{= X \setminus} \mathrm{Tr}(C)$. 
The conclusion follows directly from (5) and the definition of $T_{k}$ 
and $T_{k}'$. 
\end{proof of lemma}

\begin{theorem}
\label{theo2}
We obtain the following.
\begin{enumerate}
\def\labelenumi{(\roman{enumi})}
\item 
$F^{N1}$ \textit{is induced by} $f^{IIS}$.
\item
\textit{If} $C$ \textit{is symmetric with respect to} $X$ 
\textit{in the sense that}  $| \{ c \in C\colon c(x) =$\linebreak
True$\} | = | \{ c \in C\colon c(y) = \mathrm{True}\} |$  
\textit{for all} $x,y \in X$ \footnote{Note that this condition can be rewritten asking that there exists $k\in\mathbb{N}$ such that $|\{ c \in C\colon c(x)\}| =k$, $\forall x\in X$.}, \textit{then} $F^{N2}$ 
\textit{is induced by} $f^{S}$.
\end{enumerate}
\end{theorem}

\begin{proof of theorem}[\ref{theo2}]
\label{poftheo2}\normalfont%
\mbox{}
\begin{enumerate}
\def\labelenumi{(\roman{enumi})}
\item
  Let $x \in \max f^{IIS}\left( o_{\succsim} \right)$ and
  $y \in F^{N1}( \succsim )$. Subsequently, we obtain
  $e_{o_{\succsim}}(x) \geq e_{o_{\succsim}}(y)$. 
  By definition of $T_{k}'$, we obtain
  $T_{1}' \supseteq \cdots \supseteq T_{l}'$. Hence,
  $e_{o_{\succsim}}(x) \geq e_{o_{\succsim}}(y)$ implies that
  $x \in T_{e_{o}(x)}' \subseteq T_{e_{o}(y)}'$. By definition
  of $F^{N1}$, $y \in F^{N1}( \succsim )$ means that
  $T_{e_{o}(y)} = T_{e_{o}(y)}' \subseteq F^{N1}( \succsim )$ (the
  first equality is because of Lemma~\ref{lem2}). These imply that 
  $x \in F^{N1}( \succsim )$. Since $x,y \in X$ are arbitrary as
  long as $e_{o_{\succsim}}(x) \geq e_{o_{\succsim}}(y)$, then
  $F^{N1}( \succsim ) \supseteq \max{f^{IIS}\left( o_{\succsim} \right)}$.

\hskip2em
Assume that $e_{o_{\succsim}}(x) > e_{o_{\succsim}}(y)$. Then,
$x \in T_{e_{o_{\succsim}}(x)}' = T_{e_{o_{\succsim}}(x)}$ and
$y \notin T_{e_{o_{\succsim}(x)}}' = T_{e_{o_{\succsim}(x)}}$.
Since $T_{e_{o_{\succsim}}(x)} \neq \emptyset$, the recipe of $F^{N1}$ 
demands that $F^{N1}( \succsim ) \subseteq T_{{e_{o}}_{\succsim}(x)}$.
Hence, $y \notin F^{N1}( \succsim )$. Since $x,y \in X$ are
arbitrary as long as $e_{o_{\succsim}}(x) > e_{o_{\succsim}}(y)$, 
we obtain $\left( \max{f^{IIS}\left( o_{\succsim} \right)} \right)^{c} 
\subseteq \left( F^{N1}( \succsim ) \right)^{c}$. Therefore,
$\max f^{IIS}\left( o_{\succsim} \right) \supseteq F^{N1}( \succsim )$ and, in summary, we obtain that
$F^{N1}( \succsim ) = \max{f^{IIS}\left( o_{\succsim} \right)}$.
\item
  Under the stated condition,
\[
\begin{aligned}
F^{N2}( \succsim ) 
 & = \underset{x \in X}{\arg\max}
  \sum_{c \in C: x \in \mathrm{Tr}(c)} {s_{\succsim}^{B}(c)} \\
 & = \underset{x \in X}{\arg\max}
  \sum_{c \in C: x \in \mathrm{Tr}(c)} m_{o_{\succsim}}
  \left(
  \mathrm{Tr}(c)  \right)
  \\
 & = \underset{x \in X}{\arg\max}{\sum_{S \in \mathfrak{X}: x \in S}
  {m_{o_{\succsim}}(S)}} = \max\left( f^{S}\left( o_{\succsim} \right) \right).
\end{aligned}
\]
\end{enumerate}
\end{proof of theorem}

\section{Concluding remarks}
\label{sec4}

This study analyzes the methods for aggregating procedures over
criteria for a set of alternatives. Two promising methods,
denoted as $F^{N1}$ and $F^{N2}$ 
(Definitions ~\ref{def2} and ~\ref{def3}) are proposed in \cite{Nurmi2015}. 

Theorem~\ref{theo2} states that, for all $\succsim \in \mathcal{L}(C)^{n}$, 
we obtain,
\begin{align*}
F^{N1} (\succsim) = \max \left(f^{I I S}\left(o_{\succsim}\right)\right),
\end{align*}
and, if $C$ is symmetric with respect to $X$, then 
\begin{align*}
F^{N 2}(\succsim)=\max \left(f^S\left(o_{\succsim}\right)\right).
\end{align*}
The $\max( \cdot )$ operator transforms the weak orders
$f^{IIS}\left( o_{\succsim} \right)$ and
$f^{S}\left( o_{\succsim} \right)$ in the choice set (the set of
maximum elements with respect to the weak orders). Therefore, after
identifying $\succsim$ based on induced states of opinion
$o_{\succsim}$, these equality indicates that $F^{N1}$ and $F^{N2}$
are `equivalent' to $f^{IIS}$ and $f^{S}$, respectively.

The rationality of this interpretation depends on the plausibility of
identifying $\succsim$ with $o_{\succsim}$. As we introduce in
Definition~\ref{def6}, $o_{\succsim}$ is derived from
$\succsim = \left( \succsim_{1}, \succsim_{2},\cdots, \succsim_{n} \right)\in\mathcal{L}(C)^n$
by considering each $c \succsim_{i}d$ (for all $c,d \in C$ and $i \in N$) as
an opinion $\left( \mathrm{Tr}(c),\mathrm{Tr}(d) \right)$. The two implicit assumptions are as follows:
One assumption is that the voter label is
irrelevant \footnote{In the study of procedural choice, the anonymity of
  voters are often considered reasonable (see, e.g., \cite{Dietrich2005}).}: ``$i$
prefers $c$ to $d$'' and ``$j$ prefers $c$ to $d$'' are
counted as a single opinion $(c,d)$. Another assumption is
that criterion $c$ is identified with the set $\mathrm{Tr}(c)$ of the voting
rules satisfying $c$ (a similar assumption is made in Assumption 4 of
\cite{Suzuki2020a}). Thus, a criterion is identified using a set of all
alternatives satisfying this requirement. Based on this interpretation, the main results (Theorem~\ref{theo3}) is
the characterization of $f^{IIS}$, and together with the characterization of
$f^{S}$ in \cite{Suzuki}, we obtain the axiomatic characterization of
$F^{N1}$ and $F^{N2}$.

\section*{Acknowledgments}

Takahiro Suzuki was supported by JSPS KAKENHI Grant
Number JP21K14222. Stefano Moretti acknowledges the financial support from the ANR project THEMIS (ANR-20-CE23-0018). Michele Aleandri is member of the GNAMPA of the Istituto Nazionale di Alta Matematica (INdAM) and is supported by Project of Significant National Interest – PRIN 2022 of title “Impact of the Human Activities on the Environment and Economic Decision Making in a Heterogeneous Setting: Mathematical Models and Policy Implications”- Codice Cineca: 20223PNJ8K- CUP I53D23004320008.

\section*{Disclosure of Interests.}

The authors declare no conflict of interest. 

\bibliographystyle{plain}

\begin{thebibliography}{10}

\bibitem{Arrow1951}
KJ~Arrow.
\newblock {\em {Social choice and individual values}}.
\newblock Wiley, New York, 1951.

\bibitem{BanzhafIII1964}
J.F. {Banzhaf III}.
\newblock {Weighted voting doesn't work: A mathematical analysis}.
\newblock {\em Rutgers L. Rev.}, 19:317--344, 1964.

\bibitem{Barbera2023a}
Salvador Barber{\`{a}} and Walter Bossert.
\newblock {Opinion aggregation: Borda and Condorcet revisited}.
\newblock {\em Journal of Economic Theory}, 210:105654, 2023.

\bibitem{Barbera2023b}
Salvador Barber{\`{a}}, Walter Bossert, and Juan~D. Moreno-Ternero.
\newblock {Wine rankings and the Borda method}.
\newblock {\em Journal of Wine Economics}, 18(2):122--138, 2023.

\bibitem{Barbera2004}
Salvador Barbera and Matthew~O Jackson.
\newblock Choosing how to choose: Self-stable majority rules and constitutions.
\newblock {\em The Quarterly Journal of Economics}, 119(3):1011--1048, 2004.

\bibitem{Bernardi2019}
Giulia Bernardi, Roberto Lucchetti, and Stefano Moretti.
\newblock {Ranking objects from a preference relation over their subsets}.
\newblock {\em Social Choice and Welfare}, 52(4):589--606, 2019.

\bibitem{Brams2002}
Steven~J. Brams and Peter~C. Fishburn.
\newblock {Chapter 4 Voting procedures}.
\newblock {\em Handbook of Social Choice and Welfare}, 1:173--236, 2002.

\bibitem{Brandt2016}
Felix Brandt, Christian Geist, and Martin Strobel.
\newblock {Analyzing the Practical Relevance of Voting Paradoxes via Ehrhart Theory , Computer Simulations , and Empirical Data}.
\newblock {\em Proceedings of the 2016 International Conference on Autonomous Agents and Multiagent Systems}, pages 385--393, 2016.

\bibitem{Dietrich2005}
Franz Dietrich.
\newblock {How to reach legitimate decisions when the procedure is controversial}.
\newblock {\em Social Choice and Welfare}, 24(2):363--393, 2005.

\bibitem{Houy2004}
Nicolas Houy.
\newblock A note on the impossibility of a set of constitutions stable at different levels.
\newblock Technical report, Universit{\'e} Panth{\'e}on-Sorbonne (Paris 1), 2004.

\bibitem{Koray2006}
Semih Koray and Arkadii Slinko.
\newblock {Self-selective social choice functions}.
\newblock {\em Social Choice and Welfare}, 31(1):129--149, 2008.

\bibitem{Kultti2009}
Klaus Kultti and Paavo Miettinen.
\newblock {Stability of constitutions}.
\newblock {\em Journal of Public Economic Theory}, 11(6):891--896, 2009.

\bibitem{Laslier2011}
Jean-fran{\c{c}}ois Laslier.
\newblock {And the loser is ... Plurality Voting}.
\newblock In {\em Felsenthal, D., Machover, M. (eds) Electoral Systems. Studies in Choice and Welfare.} Springer, Berlin, Heidelberg, 2011.

\bibitem{Nurmi2015}
Hannu Nurmi.
\newblock {The choice of voting rules based on preferences over criteria}.
\newblock In Bogumi{\l} Kami{\'{n}}ski, Gregory~E. Kersten, and Tomasz Szapiro, editors, {\em Outlooks and Insights on Group Decision and Negotiation. GDN 2015. Lecture Notes in Business Information Processing, vol218}, volume 218, pages 241--252. 2015.

\bibitem{owen95}
G.~Owen.
\newblock {\em Game Theory}.
\newblock Academic Press, 1995.

\bibitem{Rae1969}
Douglas~W Rae.
\newblock {Decision rules and individual values in constitutional choice}.
\newblock {\em American Political Science Association}, 63(1):40--56, 1969.

\bibitem{Shapley1953}
Lloyd~S. Shapley.
\newblock {A value for n-person games}.
\newblock {\em Contributions to the Theory of Games}, 2(28):307--317, 1953.

\bibitem{Suzuki}
Takahiro Suzuki.
\newblock Aggregating opinions on sets of alternatives: Characterization and applications.
\newblock {\em Group Decision and Negotiation}, pages 1--23, 2025.

\bibitem{Suzuki2020a}
Takahiro Suzuki and Masahide Horita.
\newblock A characterization for procedural choice based on dichotomous preferences over criteria.
\newblock In Danielle~Costa Morais, Liping Fang, and Masahide Horita, editors, {\em Group Decision and Negotiation: A Multidisciplinary Perspective}, pages 91--103, Cham, 2020. Springer International Publishing.

\bibitem{Suzuki2023}
Takahiro Suzuki and Masahide Horita.
\newblock A society can always decide how to decide: A proof.
\newblock {\em Group Decision and Negotiation}, 32(5):987--1023, 2023.

\bibitem{Suzuki2024}
Takahiro Suzuki and Masahide Horita.
\newblock {Consistent Social Ranking Solutions}.
\newblock {\em Social Choice and Welfare}, 62:549--569, 2024.

\bibitem{Suzuki2024c}
Takahiro Suzuki and Masahide Horita.
\newblock {Sabotage‑proof social ranking solutions}.
\newblock {\em Theory and Decision}, (August), 2024.

\bibitem{Suzuki2024a}
Takahiro Suzuki, Yu~Maemura, and Masahide Horita.
\newblock {A Unified Understanding of Majority Rule, CP Majority Rule, and Their Variants.}
\newblock 2024.

\end{thebibliography}

\end{document}